\documentclass[sigconf,screen]{acmart}

\copyrightyear{2026}
\acmYear{2026}
\setcopyright{cc}
\setcctype{by}
\acmConference[ASE '26]{Proceedings of the 41st IEEE/ACM International Conference on Automated Software Engineering}{October 12--16, 2026}{Munich, Germany}
\acmBooktitle{Proceedings of the 41st IEEE/ACM International Conference on Automated Software Engineering (ASE '26), October 12--16, 2026, Munich, Germany}
\acmDOI{10.1145/3832783.3837427}
\acmISBN{979-8-4007-2882-2/2026/10}

\usepackage{nicefrac}
\usepackage{siunitx}
\usepackage{bbding}
\usepackage{pifont}
\usepackage{framed}
\usepackage{xcolor}
\definecolor{LightGray}{rgb}{0.97,0.97,0.97}
\usepackage{
  color,
  float,
  graphics,
  graphicx,
  subcaption
}
\usepackage{textcomp}
\usepackage{amsfonts}
\usepackage[ruled,linesnumbered]{algorithm2e}
\usepackage{fancyhdr,url}
\usepackage{xspace}
\usepackage{multirow}
\usepackage{balance}
\usepackage{cancel}
\usepackage[table]{xcolor}
\usepackage{colortbl}
\usepackage{pifont}
\usepackage{framed}

\newlength{\fbdesc}
\hypersetup{
    colorlinks=true,
    linkcolor=black,
    urlcolor=black,
    citecolor=black
}

\usepackage{framed}
\definecolor{shadecolor}{gray}{0.92}
\newcommand{\findingbox}[2][]{%
  \begin{shaded}%
    \setlength{\hsize}{\dimexpr\columnwidth-2\FrameSep-2\FrameRule\relax}%
    \noindent #2%
  \end{shaded}}

\usepackage{listings}

\usepackage{mathtools}
\usepackage{hyperref}

\DeclareMathAlphabet{\mathcal}{OMS}{cmsy}{m}{n}

\definecolor{tableShade}{gray}{0.92}
\definecolor{bestColor}{RGB}{144,238,144}    
\definecolor{secondColor}{RGB}{255,255,150}  
\definecolor{worstColor}{RGB}{255,182,182}   
\definecolor{highlightColor}{RGB}{230,247,255}

\DeclareGraphicsExtensions{%
    .png,.PNG,%
    .pdf,.PDF,%
    .jpg,.mps,.jpeg,.jbig2,.jb2,.JPG,.JPEG,.JBIG2,.JB2}

\newif\ifreview
\reviewtrue          

\newcommand{\authoradd}[2]{\ifreview\textcolor{#1}{#2}\else#2\fi}

\newcommand{\wbadd}[1]{\authoradd{black}{#1}}

\begin{document}

\title{How Effective Are NPM Malicious Package Detectors? A Large-Scale Empirical Study}

\author{Wenbo Guo}
\orcid{0000-0001-6655-8179}
\affiliation{%
  \institution{Nanyang Technological University}
  \city{Singapore}
  \country{Singapore}
}
\email{wenbo002@e.ntu.edu.sg}

\author{Zhongwen Chen}
\orcid{0009-0009-6446-7535}
\affiliation{%
  \institution{Sichuan University}
  \city{Chengdu}
  \country{China}
}
\email{yinglilikz@stu.scu.edu.cn}

\author{Zhengzi Xu}
\authornote{Corresponding author.}
\orcid{0000-0002-8390-7518}
\affiliation{%
  \institution{Imperial Global Singapore}
  \city{Singapore}
  \country{Singapore}
}
\email{z.xu@imperial.ac.uk}

\author{Chengwei Liu}
\orcid{0000-0003-1175-2753}
\affiliation{%
  \institution{Nankai University}
  \city{Tianjin}
  \country{China}
}
\email{chengwei.liu@nankai.edu.cn}

\author{Ming Kang}
\orcid{0009-0007-0168-1775}
\affiliation{%
  \institution{Sichuan University}
  \city{Chengdu}
  \country{China}
}
\email{kreleasem@gmail.com}

\author{Shiwen Song}
\orcid{0009-0008-7885-1135}
\affiliation{%
  \institution{Singapore Management University}
  \city{Singapore}
  \country{Singapore}
}
\email{swsong@smu.edu.sg}

\author{Chengyue Liu}
\orcid{0009-0003-7986-8004}
\affiliation{%
  \institution{Nanyang Technological University}
  \city{Singapore}
  \country{Singapore}
}
\email{CHENGYUE001@e.ntu.edu.sg}

\author{Yijia Xu}
\orcid{0000-0003-2843-4225}
\affiliation{%
  \institution{Sichuan University}
  \city{Chengdu}
  \country{China}
}
\email{xuyijia@scu.edu.cn}

\author{Weisong Sun}
\orcid{0000-0001-9236-8264}
\affiliation{%
  \institution{Nanyang Technological University}
  \city{Singapore}
  \country{Singapore}
}
\email{weisong.sun@ntu.edu.sg}

\author{Yang Liu}
\orcid{0000-0001-7300-9215}
\affiliation{%
  \institution{Nanyang Technological University}
  \city{Singapore}
  \country{Singapore}
}
\email{yangliu@ntu.edu.sg}

\renewcommand{\shortauthors}{Guo, et al.}


\begin{abstract}

The NPM ecosystem faces escalating threats from malicious
packages that exploit its open publication model. While numerous
detection tools have been proposed, they are evaluated on disparate
datasets with inconsistent settings, making cross-tool comparison
unreliable and leaving practitioners without clear guidance.

We present the first large-scale empirical study of NPM malicious
package detection, evaluating 11 tools with 16 variants on a unified
benchmark of 6,420 malicious and 7,288 benign packages annotated
with 11 behavior categories and 8 evasion techniques. Unlike prior
work, we analyze each tool’s detection mechanism to explain why tools
succeed or fail, not merely how often.
Our key findings: (1) the precision and recall a tool achieves are structurally
determined by how it resolves the ambiguity between code
capability and malicious intent, with IntelGuard reaching the best F1 at 95.98\% by
grounding its judgment in retrieved evidence and GuardDog the best among conventional tools at
93.32\%;
(2) behavioral coupling amplifies detection
signals when behaviors co-occur, raising SAP\_DT from 3.2\% to
79.3\% for the collect-and-exfiltrate chain; (3) 80.3\% of malware uses
no evasion because the ecosystem lacks mandatory pre-publication scanning; 
(4) ML degradation is driven by concept convergence rather than concept drift, since malware became simpler and every decision boundary fitted to a corpus ages with it;
(5) combination effectiveness equals complementarity minus false-positive introduction, not
paradigm diversity. Strategic combinations reach up to 97.21\% accuracy and 97.02\% F1. We release our benchmark and evaluation framework.

\end{abstract}


\maketitle

\section{Introduction}
\label{sec:introduction}




NPM serves as the cornerstone of JavaScript development, hosting over 3.57 million packages with 7.09 billion weekly downloads~\cite{npmstats, brilworks_nodejs_stats}.
Modern web applications routinely incorporate hundreds of NPM dependencies~\cite{socket_npm_retrospective, bacancy_nodejs_stats}, creating supply chain networks where a single compromised package can affect thousands of downstream projects. The NPM registry adopts an open publication model with minimal gatekeeping~\cite{ferreira2021containing, socket_npm_retrospective}, which prioritizes accessibility but also enables attackers to publish malicious packages. Reported threats include credential harvesters~\cite{securelist_lofylife}, cryptocurrency miners~\cite{duan2020towards}, and backdoor installation scripts~\cite{rptu_npm_trojan}. Attackers also exploit naming similarities through typosquatting~\cite{zimmermann2019small}, or gain access via dependency confusion and compromised maintainer accounts~\cite{zahan2022weak}. Several high-profile incidents illustrate the impact of these attacks. The \textit{event-stream} package, with 2 million weekly downloads, was compromised to steal Bitcoin wallet keys~\cite{medium_event_stream}. Attacks on \textit{ua-parser-js}~\cite{cisa_npm_malware}, \textit{coa}, and \textit{rc}~\cite{broadcom_malicious_npm} affected thousands of projects, showing that NPM has become a major vector for software supply chain attacks.

\wbadd{To counter these threats, researchers and practitioners have developed detection tools spanning four paradigms, namely static analysis of code and metadata~\cite{datadog_guarddog, microsoft_ossgadget}, dynamic analysis in a sandbox~\cite{duan2020towards, packj}, machine learning over package features~\cite{zhang2025killing, sejfia2022practical}, and large language models reasoning over code semantics, either from the model's own knowledge~\cite{zahan2024leveraging, wyss2025evaluating} or from evidence retrieved from outside it~\cite{intelguard_www2026}. Each category has shown promising detection results in their published evaluations.}


However, malicious packages remain a persistent problem despite the availability of these tools. More fundamentally, we lack understanding of \textit{why} different detection approaches succeed or fail, which prevents principled tool design and deployment. Four questions remain unanswered.

\wbadd{\textit{(Research Gap 1: Detection performance and its structural causes}.) Existing tools
exhibit precision ranging from 50\% to 99\% and recall from 44\% to 99\%. Since malicious and
benign packages invoke the same OS APIs (e.g., \textit{https.request}, \textit{child\_process.exec}), every detection tool must resolve the ambiguity between \textit{capability} and \textit{intent}. Capability is what an API makes possible, intent is the purpose it serves, and the \textit{capability-intent gap} is the distance between them that no call site reveals, since \textit{child\_process.exec} drives a build script and a
reverse shell alike. No study has traced the precision and recall a tool achieves back to how its design resolves this gap.}



\textit{(Research Gap 2: Attack-detection interaction.)} Individual tool papers report overall accuracy on their own datasets, but the relationship between specific attack characteristics (behaviors, evasion techniques, attack surface) and detection difficulty remains unclear. It is unknown whether detection failures stem from sophisticated evasion or from a mismatch between where the attack resides (e.g., package.json vs.\ JavaScript files) and what the tool analyzes.

\textit{(Research Gap 3: Temporal degradation mechanisms.)} ML-based tools trained on historical data are known to suffer performance degradation, commonly attributed to ``concept drift.'' However, the specific mechanism driving degradation in the npm ecosystem has not been established. It is unclear whether attackers develop more sophisticated evasion over time or whether changes in malware coding style, account for the observed decline.

\textit{(Research Gap 4: Combination principles.)} Practitioners are advised to deploy multiple tools for better coverage, yet no study has examined the principles governing combination effectiveness. It is unknown whether tool diversity guarantees improvement or whether certain combinations can be counterproductive.

To address these gaps, we conduct a large-scale empirical study of NPM malicious package detection tools, combining quantitative evaluation with an analysis of each tool's detection mechanism. We curate a unified benchmark of 6,420 malicious and 7,288 benign packages, annotated with 11 behavior categories and 8 evasion techniques. Our study investigates four research questions:

\begin{itemize}
\item \textbf{RQ1 (Detection Effectiveness)}: What \wbadd{is} the detection performance of current tools, and what structural mechanisms \wbadd{determine the precision and recall each tool achieves}? \textit{(Gap~1)}
\item \textbf{RQ2 (Fine-grained Behavioral Analysis)}: How do malicious behaviors, evasion techniques, and attack surfaces interact with different detection approaches? \textit{(Gap~2)}
\item \textbf{RQ3 (Temporal Evolution Analysis)}: How does detection performance evolve over time, and what mechanism drives the temporal degradation of ML-based tools? \textit{(Gap~3)}
\item \textbf{RQ4 (Tool Complementarity Analysis)}: What principles govern the effectiveness of tool combinations, and under what conditions can combination be counterproductive? \textit{(Gap~4)}
\end{itemize}

\wbadd{To answer these questions, we go beyond running tools as black boxes: we analyze the detection mechanisms of all 11 tools, including their detection rules, feature designs, thresholds, and architectural assumptions. This enables us to explain \textit{why} tools succeed or fail. We construct the benchmark from academic datasets and security advisories, build its taxonomy through LLM-assisted analysis and expert review, and evaluate 11 tools with 16 variants across packages published from 2020 to 2025.}


First, what a tool detects is decided by how it resolves the ambiguity between code
\textit{capability} and \textit{intent}, since malicious and benign packages invoke identical OS
APIs. \wbadd{IntelGuard reaches the best F1 (95.98\%) by grounding each judgment in retrieved
evidence and GuardDog the best among conventional tools (93.32\%) through taint analysis}, while
the rest span 50.41\% to 99.88\% precision and 44.41\% to 98.69\% recall. \wbadd{Our re-implementation of SocketAI~\cite{zahan2024leveraging} runs the same model as IntelGuard yet trails it by 49 points of recall, because it asks the model to judge the code alone while IntelGuard first retrieves known malware for the model to compare it
against} \textit{(RQ1)}. Second, among the 8 evasion categories, environment detection (mean
51.7\%) and anti-analysis (mean 52.7\%) are the hardest, and ML-based tools fare
worst of all, with SAP\_DT and SAP\_RF catching 5\% of anti-analysis \textit{(RQ2)}. Third, ML
degrades sharply over time, with SAP\_DT falling from 87.15\% to 39.49\% between 2021 and 2023. We
trace this to \textit{concept convergence} rather than concept drift, since attackers adopted
minimal-footprint code that is statistically indistinguishable from benign packages, so any
boundary fitted to earlier malware ages with it \textit{(RQ3)}. \wbadd{Fourth, tool combinations reach 97.21\% accuracy and 97.02\% F1, above any single tool, though the margin narrows as the best single tool improves. Combination effectiveness depends on complementarity minus false-positive introduction rather than paradigm diversity, since a high-recall, low-precision partner can degrade a strong tool (RQ4).}

The main contributions of this paper are:

\begin{enumerate}
\item \textbf{Benchmark and taxonomy}. We curate the largest NPM malware benchmark \wbadd{annotated at the behavior level} (6,420 malicious + 7,288 benign packages) with a taxonomy of 11 behavior categories and 8 evasion techniques, enabling fine-grained analysis not possible with prior datasets.
\item \textbf{Large-scale Evaluation}. We evaluate \wbadd{11} tools (\wbadd{16} variants) and, unlike prior benchmarks that treat tools as black boxes, analyze each tool’s detection mechanism to identify the structural causes behind its performance.
\item \textbf{Novel findings.} We identify five structural principles: (i) \textit{capability-intent ambiguity} as the root cause of precision-recall trade-offs; (ii) \textit{concept convergence} driving ML temporal degradation; (iii) \textit{attack surface mismatch} leaving 21\% of malware structurally undetectable by code-analysis tools; (iv) \textit{behavioral coupling amplification} boosting detection signals by up to 76 percentage points; and (v) \textit{complementarity-minus-FP-introduction} as the governing principle of tool combination effectiveness.
\item We release our dataset, evaluation framework, and all analysis scripts to facilitate future research~\cite{our_website}.
\end{enumerate}
\section{Related Work}
\label{sec:relatedwork}

\subsection{NPM Supply Chain Security}

When developers run \texttt{npm install}, npm resolves dependencies and automatically executes the
\texttt{preinstall}, \texttt{install}, and \texttt{postinstall} lifecycle scripts with full user
privileges before any explicit code invocation~\cite{zahan2022weak, ferreira2021containing}, after
which packages may freely access file systems, network resources, and environment
variables~\cite{duan2020towards, ohm2020backstabber, ladisa2023sok}. \wbadd{Only 2.2\% of packages
declare such scripts~\cite{zahan2022weak}, yet what those scripts execute propagates to every
dependent project, since an average package implicitly trusts 79 third-party
packages}~\cite{zimmermann2019small}. Attackers reach this surface \wbadd{either by inducing a
victim to install the wrong package or by subverting one the victim already trusts. Name confusion
spans the 13 mechanisms Neupane et al.\ separate across 1{,}232 documented attacks, of which
typosquatting is only one}~\cite{neupane2023beyond, taylor2020defending,
vu2020typosquatting}\wbadd{, dependency confusion exploits name resolution by publishing public
packages that match private internal names}~\cite{ladisa2023sok, abdalkareem2020impact,
cao2022towards, latendresse2022not}\wbadd{, and account takeover requires no new
package}~\cite{garrett2019detecting, eslint_postmortem_2018, zimmermann2019small}.

\wbadd{Prior npm work either characterizes the attacks without measuring detectors, or measures
detectors without explaining them. Backstabber's Knife Collection organizes malicious packages into
attack trees~\cite{ohm2020backstabber}, Ladisa et al.\ generalize such observations into a taxonomy
of attack vectors~\cite{ladisa2023sok}, and Zhou et al.\ relate a large corpus of them through a
knowledge graph~\cite{zhou2025analysis}, yet none of the three reports how a detector responds to
the attacks it catalogues. Zahan et al.\ compare two GPT models with CodeQL~\cite{zahan2024leveraging},
Wyss et al.\ test LLMs on version diffs~\cite{wyss2025evaluating}, and ProfMal~\cite{profmal_ase2025}
is evaluated against five detectors, but each reports aggregate scores rather than the mechanism
that produces them. We evaluate 11 tools across 16 variants covering all four paradigms on identical
inputs, label each package with 11 behavior and 8 evasion categories, track detection from 2011 to
2025, and analyze each tool’s detection mechanism to explain why it succeeds or fails.}

\subsection{NPM Malicious Package Detection}

\begin{table}[t]
\centering
\caption{Overview of Studied npm Malicious Package Detection Tools}
\label{tab:malicious_npm_detection_tools}
\scriptsize
\setlength{\tabcolsep}{4pt}
\renewcommand{\arraystretch}{1.1}
\resizebox{\columnwidth}{!}{%
\begin{tabular}{@{}lllllll@{}}
\toprule
\textbf{Type} & \textbf{Name} & \textbf{Year} & \textbf{Target} & \textbf{Feature} & \textbf{Classifier} & \textbf{Avail.} \\
\midrule
\multirow{6}{*}{\textbf{Static-Based}} 
& LastJSMile~\cite{scalco2022feasibility} & 2022 & Phantom files & Code diff, Metadata & RegEx & \ding{55} \\
& OSSGadget~\cite{microsoft_ossgadget} & 2024 & Full package & Code, Metadata & Rule-based & \ding{52} \\
& GuardDog~\cite{datadog_guarddog} & 2024 & Full package & Code, Metadata & Semgrep & \ding{52} \\
& diff-CodeQL~\cite{diff-CodeQL} & 2023 & Full package & Code diff & CodeQL & \ding{55} \\
& GENIE~\cite{GENIE} & 2024 & Full package & Code & CodeQL & \ding{52} \\
& MalWuKong~\cite{li2023malwukong} & 2023 & Full package & Code, API call & CodeQL & \ding{55} \\
\midrule
\multirow{2}{*}{\textbf{Dynamic-Based}} 
& MalOSS~\cite{duan2020towards} & 2021 & Source, binaries & API call, Dynamic & Rule-based & \ding{52} \\
& Packj~\cite{packj} & 2023 & Full package & API call, Dynamic & Rule-based & \ding{52} \\
\midrule
\multirow{10}{*}{\textbf{ML-Based}} 
& Cerebro~\cite{zhang2025killing} & 2023 & Full package & AST & BERT & \ding{52}$^{*}$ \\
& SAP~\cite{ladisa2023feasibility} & 2023 & Full package & Statistical & DT, RF, XGB & \ding{52} \\
& AMALFI~\cite{sejfia2022practical} & 2022 & Full package & API call, Metadata, Update & DT, NB, SVM & \ding{55} \\
& DONAPI~\cite{huang2024donapi} & 2024 & Full package & AST, API sequence, Dynamic & RF & \ding{55} \\
& Maltracker~\cite{yu2024maltracker} & 2024 & Full package & Code, Graph & DT, RF, XGB & \ding{55} \\
& Ohm et al.~\cite{ohm2022feasibility} & 2022 & Full package & Metadata, Code & SVM, MLP, RF & \ding{55} \\
& MeMPtec~\cite{halder2024malicious} & 2024 & Metadata & Metadata & SVM, GLM, GBM, DRF, DL & \ding{55} \\
& Maldet~\cite{zhang2024maldet} & 2024 & Full package & Statistical, API call, Metadata & DT, RF, NB, SVM & \ding{55} \\
& SpiderScan~\cite{huang2024spiderscan} & 2024 & Full package & Code, API call, Dynamic & RF & \ding{55} \\
& MalPacDetector~\cite{11037372} & 2025 & Full package & AST, Feature Set & RF, SVM, NB, MLP & \ding{52} \\
& \wbadd{ProfMal~\cite{profmal_ase2025}} & \wbadd{2025} & \wbadd{Full package} & \wbadd{Behavior graph, Dynamic} & \wbadd{Graph classifier} & \wbadd{\ding{52}} \\
& \wbadd{EMPHunter~\cite{emphunter_tse2025}} & \wbadd{2025} & \wbadd{Install scripts} & \wbadd{Script similarity} & \wbadd{Clustering} & \wbadd{\ding{52}} \\
\midrule
\textbf{LLM-Based} 
& SocketAI~\cite{zahan2024leveraging} & 2025 & Full package & Code Content & LLMs & \ding{52}$^{*}$ \\
& \wbadd{IntelGuard~\cite{intelguard_www2026}} & \wbadd{2026} & \wbadd{Full package} & \wbadd{Code, Threat intel.} & \wbadd{LLMs, RAG} & \wbadd{\ding{52}} \\
\bottomrule
\end{tabular}%
}
\vspace{2pt}

\scriptsize
\textit{Note:} Avail.=Available. \ding{52}=Yes, \ding{55}=No, \ding{52}$^{*}$=Code not public (Cerebro was evaluated with help from its authors; SocketAI was re-implemented from its paper~\cite{zahan2024leveraging}). DT=Decision Tree, RF=Random Forest, XGB=XGBoost, NB=Naive Bayes, SVM=Support Vector Machine, MLP=Multi-Layer Perceptron.
\end{table}

Detection tools differ in the evidence they use to resolve the capability-intent ambiguity, and
Table~\ref{tab:malicious_npm_detection_tools} groups them into four paradigms.
(1) \textit{Static-based} methods analyze the code without executing it.
OSSGadget~\cite{microsoft_ossgadget} and GuardDog~\cite{datadog_guarddog} match rules and Semgrep
patterns against code and metadata, LastJSMile~\cite{scalco2022feasibility} matches regular
expressions against phantom files, and GENIE~\cite{GENIE}, MalWuKong~\cite{li2023malwukong}, and
diff-CodeQL~\cite{diff-CodeQL} track information flow with CodeQL to trace a sensitive call back to
its source. Their evidence is confined to what the text itself expresses.
(2) \textit{Dynamic-based} methods execute the package and observe its runtime behavior, with
MalOSS~\cite{duan2020towards} tracing sensitive API invocations and Packj~\cite{packj} adding a sandbox to
its metadata checks. Execution yields evidence of behavior rather than of appearance, at the cost
of running untrusted code.
(3) \textit{ML-based} methods fit a decision boundary to a corpus and differ in the features they
fit it to, from statistical properties~\cite{ladisa2023feasibility} and ASTs~\cite{zhang2025killing,
11037372} to call graphs~\cite{yu2024maltracker}, API sequences~\cite{huang2024donapi,
huang2024spiderscan}, and metadata~\cite{halder2024malicious, ohm2022feasibility, zhang2024maldet,
sejfia2022practical}. \wbadd{ProfMal~\cite{profmal_ase2025} builds a behavior graph for each
package, resolving by execution the calls that static analysis cannot, and classifies the completed
graph. EMPHunter~\cite{emphunter_tse2025} requires no malicious examples, clustering the
installation scripts of new packages and ranking the outliers. The corpus therefore bounds what
they can distinguish.
(4) \textit{LLM-based} methods prompt a model to interpret the code.
SocketAI~\cite{zahan2024leveraging} queries GPT over three rounds and directs the model to
challenge its own conclusions, and Wyss et al.~\cite{wyss2025evaluating} apply the same design to
version diffs,} \wbadd{whereas IntelGuard~\cite{intelguard_www2026} retrieves precedent, distilling
over 8,000 threat intelligence reports into a knowledge base and asking the model whether the
code's behavior matches the package's stated purpose}. \wbadd{Their evidence is either the model's own
knowledge or what a retrieval step supplies.
Each paradigm has advanced on its own dataset, experimental setting, and choice of baselines.
We compare them on one benchmark and examine why each succeeds or fails.}

\section{Evaluation Framework}
\label{sec:evaluation}

\subsection{Dataset Construction}
\label{sec:dataset}

\subsubsection{Malicious Dataset Construction}

To construct a representative malicious dataset, we systematically collect NPM malware packages from two categories of sources: academic datasets and public security advisories.

\textbf{Academic Datasets.} Table~\ref{tab:malicious_npm_datasets} lists eleven released datasets, \wbadd{of which five make their NPM source downloadable together with machine-readable version and year metadata}. We collect all five: BKC~\cite{ohm2020backstabber} (2,113 versions), DONAPI~\cite{huang2024donapi} (1,159), DataDog~\cite{datadog_malicious_software_packages} (935), MalOSS~\cite{duan2020towards} (567), and Maltracker~\cite{yu2024maltracker} (230), spanning 2020 to 2024.
\textbf{Security Advisories.} OSV and Snyk record reported malicious NPM packages by name and version, but the registry removes those packages once they are disclosed. Since removed packages often remain in downstream mirrors~\cite{guo2023empirical}, we take the names and versions from the advisories and retrieve the source from those mirrors, which recovers 4,547 versions the registry no longer serves.

\subsubsection{Benign Dataset Construction}

For benign packages, we collected 7,288 packages from the official NPM registry by monthly download count, taking the latest version of each. Popularity drives the selection for three reasons. Popular packages carry larger and more varied codebases and trigger significantly higher false-positive rates than randomly sampled ones~\cite{vu2023bad, darkreading2023false}, which makes the precision we measure on them a conservative estimate. They are also the least likely to carry undiscovered malware, since a large user base and community scrutiny find it quickly. And they are what developers actually deploy, so a detector's performance on them reflects the setting it faces in practice. We verified against OSV and Snyk that none of the 7,288 is a known malicious package.


\begin{table}[t]
\centering
\caption{Dataset Sources}
\label{tab:malicious_npm_datasets}
\scriptsize
\setlength{\tabcolsep}{4pt}
\renewcommand{\arraystretch}{1.0}
\resizebox{\columnwidth}{!}{%
\begin{tabular}{@{}lccccc|lccccc@{}}
\toprule
\textbf{Paper} & \textbf{Count} & \textbf{Vers.} & \textbf{Year} & \textbf{Upd.} & \textbf{Avail.} & 
\textbf{Paper} & \textbf{Count} & \textbf{Vers.} & \textbf{Year} & \textbf{Upd.} & \textbf{Avail.} \\ 
\midrule
Cerebro~\cite{zhang2025killing} & 1,789 & 1,789 & 2023 & \ding{55} & \ding{55} &
MalOSS~\cite{duan2020towards} & 567 & 567 & 2020 & \ding{55} & \ding{52} \\
SocketAI~\cite{zahan2024leveraging} & 2,180 & 2,776 & 2024 & \ding{55} & \ding{55} &
BKC~\cite{ohm2020backstabber} & 2,113 & 2,113 & 2020 & \ding{52} & \ding{52} \\
AMALFI~\cite{sejfia2022practical} & 643 & 643 & 2022 & \ding{55} & \ding{55} &
Scalco et al.~\cite{scalco2022feasibility} & 119 & 119 & 2022 & \ding{55} & \ding{55} \\
SAP~\cite{ladisa2023feasibility} & 102 & 102 & 2023 & \ding{55} & \ding{55} &
DataDog~\cite{datadog_malicious_software_packages} & 821 & 935 & 2024 & \ding{52} & \ding{52} \\
DONAPI~\cite{huang2024donapi} & 1,159 & 1,159 & 2024 & \ding{55} & \ding{52} &
Maltracker~\cite{yu2024maltracker} & 230 & 230 & 2024 & \ding{55} & \ding{52} \\
SpiderScan~\cite{huang2024spiderscan} & 364 & 364 & 2024 & \ding{55} & \ding{55} &
\textbf{Ours} & \textbf{4,286} & \textbf{6,420} & \textbf{2025} & \textbf{\ding{52}} & \textbf{\ding{52}} \\
\bottomrule
\end{tabular}%
}
\vspace{2pt}
\scriptsize
\textit{Note:} Vers.=Versions, Upd.=Update, Avail.=Available. \ding{52}=Yes, \ding{55}=No.
\end{table}

\subsection{Dataset Preprocess}

\textbf{Deduplication.} We remove 262 duplicate entries from the original 9,551 malicious versions, leaving 9,289. \textbf{Placeholder Removal.} When NPM removes a malicious package, it replaces the content with a placeholder version marked \texttt{0.0.1-security}~\cite{phylum_npm_security_holding}, and we remove 1,502 such versions, leaving 7,787. \textbf{Manual Verification.} Source datasets may contain false positives due to mislabeling or outdated annotations. To ensure ground truth quality, three security experts holding PhDs in software supply chain security independently review each package. \wbadd{Since malicious behavior is almost always tied to sensitive resources such as processes, memory, and data~\cite{zhao2024models}, we locate those operations with GuardDog and OSSGadget, and each expert judges intent at those locations rather than reading the whole package.} A package is labeled as malicious only when at least two experts agree \wbadd{that it performs a recognizable malicious action}. Across all 7,787 reviewed packages, the three-way consensus rate is 97.4\%, and the average pairwise Cohen's $\kappa$ is 0.78. This process removes 1,367 false positives, requiring approximately 287 expert-hours in total. After preprocessing, we obtain a curated dataset of 6,420 malicious versions across 4,286 unique packages. Table~\ref{tab:malicious_npm_datasets} summarizes the final dataset statistics and sources.

\subsection{Malicious Behavior Taxonomy}
\label{sec:taxonomy}

To enable fine-grained analysis of detection capabilities, we construct a taxonomy of malicious behaviors through three steps: malicious code localization, context extraction, and behavior clustering.

\textbf{Malicious Code Localization.} Manually locating malicious code
across 6,420 packages is \wbadd{challenging}, so we automate this step using
GuardDog and OSSGadget, two complementary static analysis tools.
GuardDog's taint-based rules identify 10,059 locations
(1.57 per package), focusing on data exfiltration and command execution
patterns. OSSGadget's regex rules identify 27,350 locations (4.26
per package) with broader coverage including encoded strings and dynamic
code evaluation. \wbadd{Neither tool suffices alone, since GuardDog reports no match for 647
packages (10.08\%) and OSSGadget none for 542 (8.44\%), but} together they detect 6,352 packages,
leaving only 68 packages (1.06\%) requiring manual analysis. For locations flagged
by both tools, we retain one instance.

\textbf{Code Context Extraction.} \wbadd{Each GuardDog or OSSGadget report gives a file path, a
line number, the rule that fired, and the flagged code itself, which is} only 1-3 lines and
insufficient to determine intent, since the definitions and call chains that give those lines
meaning often lie elsewhere in the file. We therefore use \wbadd{GPT-4.1 with temperature 0 and
top-p 0.3} to recover that context. \wbadd{The input is that report paired with the entire source
file it points into, and the output is a JSON object whose \texttt{malicious\_code} field holds
the extracted context.} We prompt the LLM to extract the variable definitions and assignments the
flagged line depends on, the function call chains invoking the malicious operations, and the
control flow structures determining execution conditions. The LLM only expands context and never
decides whether code is malicious. Ground-truth labels come entirely from the dataset. For the 68
packages evading both tools, security experts perform manual extraction following the same method.

We extract 17,285 code contexts from all detections and obtain 5,699 unique contexts after deduplication. The detection count exceeds the package count because a single package often contains multiple locations, such as an install script in \texttt{package.json} and a payload in a JavaScript file. The unique count is lower than the package count because typosquatting campaigns reuse identical payloads across many packages with only the name changed.

\textbf{Behavior Summary and Clustering.} 
From each extracted code context, we prompt the LLM to generate a structured behavioral summary
that captures (i) the triggering condition/entry point, (ii) the main malicious actions (e.g.,
collection, execution, exfiltration), and (iii) the key targets or artifacts involved (e.g.,
tokens, environment variables, files, network endpoints), yielding 5,699 summaries, one per
unique context. We encode each summary using Sentence-BERT
(all-mpnet-base-v2)~\cite{reimers2019sentence}, apply UMAP for dimensionality reduction, and
cluster the embeddings using K-Means~\cite{mcqueen1967some}. 
To determine the number of clusters, we sweep $K$ over a candidate range and select the value
that jointly optimizes clustering quality (e.g., silhouette score) and stability across random
seeds, while yielding semantically coherent and interpretable groups upon manual inspection. Three domain experts (the same annotators as in Section~\ref{sec:dataset})
independently reviewed the clusters against the original code contexts, assigned a label to each
cluster, and proposed merges for functionally overlapping groups. Initial pairwise label
agreement reached 84.2\%, after which the experts resolved disagreements through discussion. \wbadd{Eleven categories survive this consolidation, so their number comes from the data rather
than from a taxonomy fixed in advance.} The resulting behaviors include data exfiltration,
arbitrary command execution, credential theft, environment reconnaissance, and persistence
installation.

\wbadd{\textbf{Context Extraction Validation.} To evaluate the accuracy of our LLM-based code
context extraction, we randomly sample 500 extracted contexts and have three security experts
independently verify whether each context correctly captures the complete malicious code segment
with its surrounding dependencies. A context is marked as correct only when at least two experts
agree. The validation achieves 97.2\% accuracy (486/500 correct) with a 97.4\% consensus rate
among experts, confirming the reliability of our extraction approach.}

\wbadd{\textbf{Behavior Summary Validation.} To evaluate the accuracy of LLM-generated behavior
summaries, we randomly sample 500 summaries and have three security experts independently verify
whether each summary correctly describes the malicious behavior. A summary is marked as correct
when at least two experts agree. The validation achieves 97.0\% accuracy (485/500 correct) with
97.6\% consensus rate and Fleiss' Kappa of 0.74, indicating substantial inter-rater agreement.
Among the 15 incorrect samples, 8 cases miss important behaviors that the code actually has, 4
cases claim behaviors that the code does not have, and 3 cases describe behaviors that do not
match actual code behavior.}

\begin{figure}[t]
    \centering
    \includegraphics[width=0.7\linewidth]{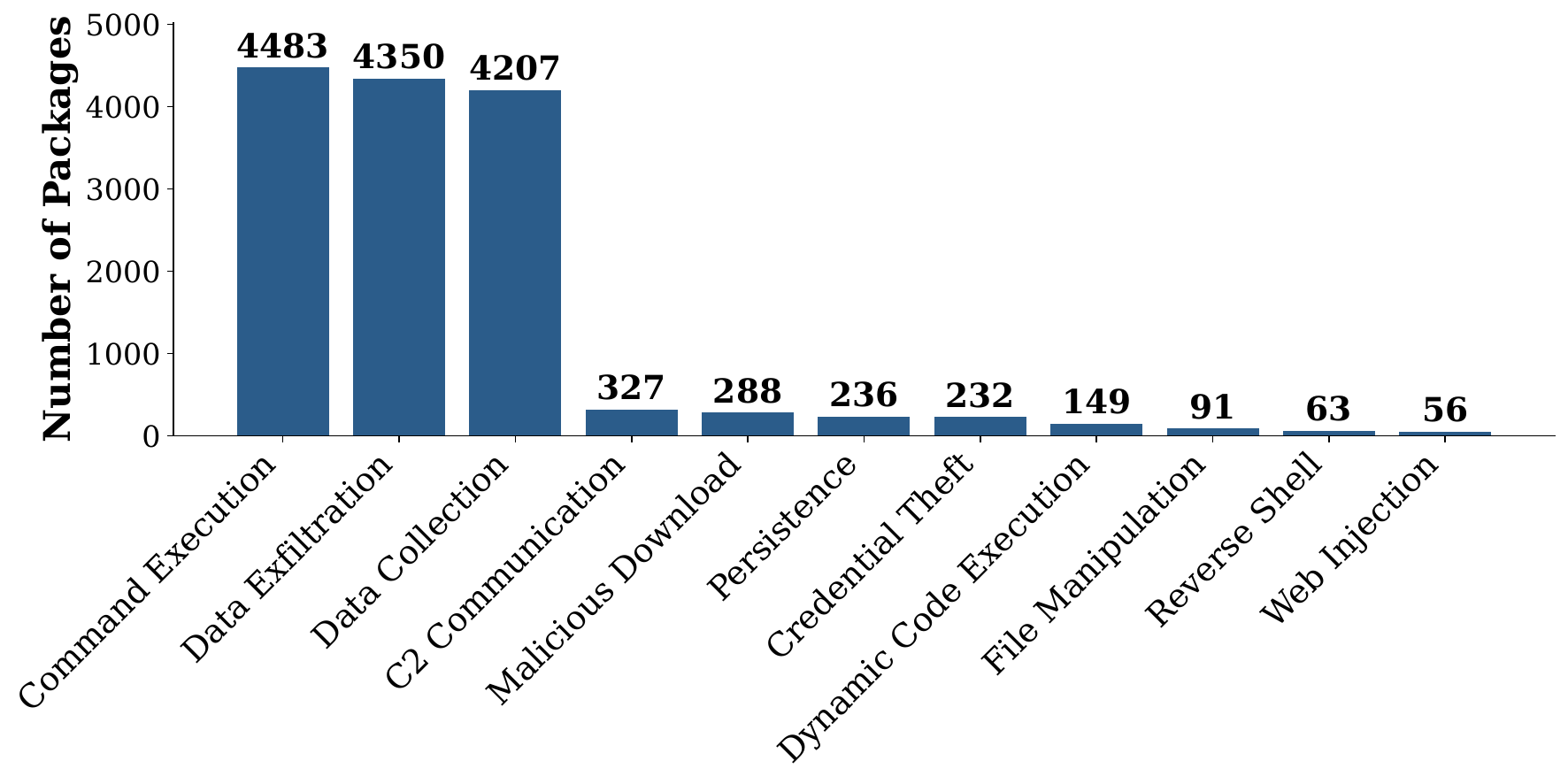}
    \caption{Behavior Distribution}
    \label{fig:behavior_distribution}
\end{figure}

\textbf{Taxonomy Analysis.}
Figure~\ref{fig:behavior_distribution} shows the distribution of 11 malicious behavior categories
across our dataset. The three most prevalent behaviors are command execution (4,483 packages),
data exfiltration (4,350), and data collection (4,207), which together form the core attack
pattern in NPM malware: collecting sensitive information and transmitting it to
attacker-controlled servers via shell commands. Other notable behaviors include C2 communication
(327), malicious download (288), and persistence (236), which indicate more sophisticated attack
chains involving remote control and long-term access. Less frequent but critical behaviors such as
credential theft (232), dynamic code execution (149), and reverse shell (63) reflect targeted
attack objectives including runtime payload delivery and interactive remote access.

Malicious packages typically combine multiple behaviors, with a mean of 2.4 and a median of 3 per
package. Across all pairs of the 11 categories, data collection and data exfiltration co-occur
most frequently in 4,144 packages, followed by command execution with data exfiltration (3,020)
and data collection (2,903), confirming a dominant \textit{collect-and-exfiltrate} pattern in
which shell commands serve as the shared mechanism for gathering and transmitting sensitive
information. Rarer pairings such as credential theft with data exfiltration and command execution
with persistence (158) point to the same long-term-access chains.

\begin{table*}[t]
\centering
\caption{Taxonomy of Malicious Behavior Categories}
\label{tab:malicious_behaviors}
\scriptsize
\setlength{\tabcolsep}{4pt}
\begin{tabular}{@{}p{2.2cm}p{\fbdesc}|p{2.2cm}p{\fbdesc}@{}}
\toprule
\textbf{Category} & \textbf{Description} & \textbf{Category} & \textbf{Description} \\
\midrule
Command Execution & Executing OS-level shell commands via child process APIs (e.g., \texttt{exec}, \texttt{spawn}) &
Data Exfiltration & Transmitting collected data to attacker-controlled servers (e.g., HTTPS POST, DNS queries) \\
Data Collection & Gathering host and environment information (e.g., hostname, username, network interfaces) &
C2 Communication & Establishing persistent bidirectional connections to remote servers for receiving commands \\
Malicious Download & Fetching and executing remote payloads or platform-specific binaries &
Persistence & Maintaining long-term access by injecting code into existing applications (e.g., Discord modules) or installing backdoors \\
Credential Theft & Stealing authentication materials (e.g., SSH keys, API tokens, browser cookies, crypto wallets) &
Dynamic Code Exec. & Constructing and executing code at runtime via \texttt{eval} or deobfuscation \\
File Manipulation & Unauthorized file system operations such as writing payloads to disk or deleting evidence &
Reverse Shell & Opening interactive shell connections to remote servers \\
Web Injection & Injecting malicious content into web pages (e.g., iframe injection, formjacking, phishing redirects) & & \\
\bottomrule
\end{tabular}
\end{table*}

\subsection{Detection Tools Selection}

\wbadd{\textbf{Tool Identification.} To collect candidate tools, we start from two established surveys of open-source supply-chain attacks, Backstabber's Knife Collection~\cite{ohm2020backstabber} and the SoK of Ladisa et al.~\cite{ladisa2023sok}. From these two papers we follow both their references and the papers that cite them, and we further search Google Scholar, the ACM Digital Library, and IEEE Xplore with terms such as \textit{npm}, \textit{malicious package}, \textit{supply chain}, and \textit{malware detection}, repeating this expansion until no new tool appears. The search yields 494 unique papers. Among them, 79 concern the npm or JavaScript ecosystem, and 61 of those propose or apply a detector for malicious packages. The remaining papers study other ecosystems, measure the threat without building a detector, or address different problems such as vulnerability or hallucination detection.}

We select tools \wbadd{from these candidates} based on four criteria: (1)~\textbf{Diversity}: covering static, dynamic, ML, and LLM approaches to ensure comprehensive coverage; 
(2)~\textbf{Impact}: prioritizing tools \wbadd{published at top venues, highly cited, or} developed \wbadd{and deployed} by major \wbadd{vendors} such as  Microsoft and Datadog; (3)~\textbf{Recency}: published or significantly 
updated within 2020--\wbadd{2026}; (4)~\textbf{Availability}: publicly accessible with API or CLI interfaces to ensure reproducibility.

We collected \wbadd{11} NPM malicious package detection tools representing \wbadd{16} distinct detection variants from academic research and industry practice. The evaluation includes: (1) \textbf{Static analysis tools}: GuardDog, OSSGadget, and GENIE, which analyze source code and metadata without execution; (2) \textbf{Dynamic analysis tool}: Packj, configured for both static analysis and dynamic trace modes; (3) \textbf{Machine learning tools}: SAP implemented with three classifiers (XGBoost, Random Forest, and Decision Tree), MalPacDetector implemented with three classifiers (MLP, Naive Bayes, and SVM), Cerebro which employs BERT to learn semantic patterns from behavior sequences\wbadd{, ProfMal which classifies behavior graphs built from combined static and dynamic analysis, and EMPHunter which clusters installation scripts and ranks outliers without requiring malicious examples}. Although MalPacDetector uses LLMs, we classify it as ML-based as its features are ultimately constructed and classified by conventional ML models; (4) \textbf{LLM-based tool\wbadd{s}}: SocketAI, the method proposed by Zahan et al.~\cite{zahan2024leveraging}, which we re-implement from the prompts and the three-stage workflow in the paper (Initial Report generation, Critical Reports analysis, and Final Report synthesis)\wbadd{, and IntelGuard, which retrieves known-malicious examples from a threat-intelligence knowledge base to ground the judgment of the LLM}. To manage computational expenses, we configured SocketAI and IntelGuard with GPT-4.1 mini as the underlying language model, utilizing the Final Report output for malicious package classification. Table~\ref{tab:malicious_npm_detection_tools} summarizes the detection tools.


\subsection{Experimental Setup}
\label{subsec:experiment_setup}

\textbf{Environment.} We evaluate detection tools on our unified dataset of 6,420 malicious package versions and 7,288 benign packages constructed as described in Section~\ref{sec:dataset}. Each tool processes packages using default configuration parameters to ensure reproducibility and reflect real-world deployment scenarios. Static analysis tools (GuardDog, GENIE, OSSGadget, Packj\_static), dynamic analysis tool (Packj\_trace), machine learning-based tools (SAP variants, MalPacDetector variants, Cerebro), and LLM-based tool (SocketAI) process packages directly through their respective detection pipelines. All experiments are conducted on Ubuntu 22.04.5 LTS systems equipped with 32GB RAM and Intel Xeon processors.

\section{Empirical Results}
\label{sec:empirical}

\subsection{RQ1: Detection Effectiveness}



\begin{table}[!ht]
\centering
\scriptsize
\caption{Performance Comparison of Detectors on Datasets}
\label{tab:F1_perform}
\renewcommand{\arraystretch}{0.85}
\begin{tabular}{llcccc}
\toprule
\textbf{Type} & \textbf{Detector} & \textbf{Accuracy} & \textbf{Precision} & \textbf{Recall} & \textbf{F1-Score} \\
\midrule
\multirow{4}{*}{\textbf{Static-Based}} 
& OSSGadget & \cellcolor{red!20}53.87\% & \cellcolor{red!20}50.41\% & 91.56\% & 65.02\% \\
& GuardDog & \cellcolor{yellow!30}93.97\% & 96.99\% & 89.92\% & \cellcolor{yellow!30}93.32\% \\
& GENIE & 74.72\% & \cellcolor{yellow!30}99.76\% & 46.14\% & 63.10\% \\
& Packj\_static & 53.91\% & \cellcolor{red!20}50.41\% & \cellcolor{green!30}98.69\% & 66.73\% \\
\midrule
\textbf{Dynamic-Based} 
& Packj\_trace & 63.91\% & 56.82\% & \cellcolor{yellow!30}95.47\% & 71.24\% \\
\midrule
\multirow{8}{*}{\textbf{ML-Based}}
& SAP\_DT & 83.86\% & \cellcolor{green!30}99.88\% & 65.62\% & 79.21\% \\
& SAP\_RF & 84.16\% & \cellcolor{green!30}99.88\% & 66.26\% & 79.67\% \\
& SAP\_XGB & 89.99\% & 94.47\% & 83.52\% & 88.66\% \\
& MalPacDetector\_MLP & 84.90\% & 97.03\% & 73.00\% & 83.32\% \\
& MalPacDetector\_NB & 85.15\% & 97.74\% & 72.92\% & 83.53\% \\
& MalPacDetector\_SVM & 85.70\% & 98.00\% & 73.81\% & 84.20\% \\
& Cerebro & 81.52\% & 99.23\% & 58.80\% & 73.85\% \\
& \wbadd{ProfMal} & \wbadd{88.52\%} & \wbadd{97.55\%} & \wbadd{77.45\%} & \wbadd{86.34\%} \\
\midrule
\multirow{2}{*}{\textbf{LLM-Based}}
& SocketAI & 73.58\% & 98.21\% & \cellcolor{red!20}44.41\% & \cellcolor{red!20}61.16\% \\
& \wbadd{IntelGuard} & \cellcolor{green!30}\wbadd{96.32\%} & \wbadd{98.24\%} & \wbadd{93.82\%} & \cellcolor{green!30}\wbadd{95.98\%} \\
\bottomrule
\end{tabular}
\end{table}

\begin{table}[t]
\centering
\scriptsize
\caption{\wbadd{EMPHunter Ranking Performance on Malicious Versions Declaring Install Hooks}}
\label{tab:emphunter_perform}
\renewcommand{\arraystretch}{0.9}
\wbadd{\begin{tabular}{lcccc}
\toprule
\textbf{Metric} & \textbf{Top-1} & \textbf{Top-3} & \textbf{Top-5} & \textbf{Top-10} \\
\midrule
Recall    & 48.57\% & 74.89\% & 76.03\% & 80.46\% \\
FPR       & 0.46\%  & 2.03\%  & 3.82\%  & 8.28\%  \\
Precision & 48.57\% & 24.96\% & 15.21\% & 8.05\%  \\
mAP@K     & 48.57\% & 61.53\% & 61.79\% & 62.35\% \\
\bottomrule
\end{tabular}}
\vspace{2pt}

\scriptsize
\wbadd{\textit{Note:} Top-K considers the K highest-ranked candidates that EMPHunter reports, and a malicious version counts as recalled when it appears among them. mAP@K averages the reciprocal rank $1/r$ of each malicious version, where the term is zero when the version falls outside the top K.}
\end{table}


\textbf{Results \& Analysis.} Table~\ref{tab:F1_perform} presents the 
detection performance of all evaluated tools. Detection effectiveness varies 
dramatically across paradigms. Among static-based tools, GuardDog achieves 
the best overall balance with an F1-score of 93.32\%, while GENIE sacrifices 
recall for near-perfect precision and Packj\_static does the opposite, 
generating 6,234 false positives to recover 98.69\% of malicious packages. 
Dynamic analysis improves recall to 95.47\% but compounds the false positive 
problem, with Packj\_trace producing 4,656 FP. 
ML-based tools exhibit a consistent precision-recall trade-off: SAP\_DT and SAP\_RF attain near-perfect precision yet miss roughly one-third of malicious packages, whereas SAP\_XGB\wbadd{,} the MalPacDetector variants\wbadd{, and ProfMal} offer a more balanced trade-off with F1-scores 
ranging from 83.32\% to 88.66\%. \wbadd{The sharpest contrast appears within the LLM paradigm.} SocketAI achieves only 44.41\% recall, missing 3,569 packages\wbadd{, whereas IntelGuard attains the highest F1-score of all evaluated tools (95.98\%) by grounding its judgments in retrieved threat intelligence knowledge base. EMPHunter ranks candidate packages rather than classifying them and is therefore reported separately (Table~\ref{tab:emphunter_perform})}.


We analyze the detection logic of each tool to understand why these performance differences exist.

\textit{Capability-end tools} flag any invocation of sensitive APIs regardless
of context. Packj\_static maps 56 JavaScript APIs to 10 permission categories
and flags any matching package, achieving 98.69\% recall but a
signal-to-noise ratio of 1:71. OSSGadget's false positives arise largely from
matching keywords such as \texttt{curl} and \texttt{wget} in README files
and comments rather than executable code, a structural flaw no rule update
can fully resolve.

\textit{Intent-end tools} impose stricter evidence requirements at the cost
of recall. GENIE's \texttt{theft-os} query demands $\geq$3 unique OS data
sources flowing to the same HTTP sink, creating a step-function boundary:
multi-source exfiltration is caught with near-perfect precision, while
single- or dual-source attacks are missed entirely, and non-HTTP exfiltration
channels are excluded by definition. SAP\_DT/RF achieve 99.88\% precision
because their learned boundaries are so narrow that only patterns closely
matching the 102-sample training set are flagged.


\textit{GuardDog} sits between these two extremes by using taint analysis
that requires end-to-end data flow from sensitive sources to network sinks as
partial evidence of malicious intent, combined with an allowlist of known
legitimate hook patterns to suppress false positives. Notably, just two of
its 22 rules account for over 90\% of detections: \texttt{npm-install-script}
targeting lifecycle script abuse and \texttt{shady-links} targeting suspicious
outbound URLs, revealing that NPM malware is homogeneous in its
attack entry point.

\textit{Among ML-based tools}, feature quality matters more than classifier
choice. MalPacDetector's 24 security-oriented binary features yield consistent
accuracy across all three classifiers, while SAP's 140 features are
structurally compromised: 65\% are file-extension counts with no security
semantics, and its two ostensibly discriminative features turn out to be
reverse signals, with dangerous token counts and Base64 chunk counts both
scoring higher in benign packages than in malicious ones. \wbadd{ProfMal confirms the same principle from the opposite direction. Its behavior 
graphs, built by object-sensitive static analysis and completed through dynamic 
execution, are the richest feature representation in this group and yield its 
highest precision (97.55\%). Yet even graph evidence trades recall away (77.45\%, 
1,448 missed versions), because attacks confined to install scripts leave little 
for a graph to capture.}

\textit{LLM-Based Tools.} SocketAI's low recall has three structural causes:
(1) its Critical Reports stage explicitly instructs the LLM to challenge
prior judgments about malicious behavior, systematically downgrading initially
correct detections; (2) LLMs produce conservative scores, placing genuine
malware below the 0.5 classification threshold; and (3) a 20-file-per-package
limit causes payloads in deep directory structures to be skipped entirely.
At 146 seconds per package, it is also costly for large-scale scanning. \wbadd{IntelGuard avoids these failure modes by design. It slices package 
files, retrieves similar known-malicious examples from a knowledge base 
distilled from over 8,000 threat intelligence reports, and lets the LLM judge 
each slice against these precedents. Anchoring the judgment to retrieved 
evidence, rather than asking the model to challenge itself, preserves 93.82\% 
recall at 98.24\% precision (108 false positives). The two tools even run the 
same underlying model, GPT-4.1 mini, so the 49 point recall gap between 
them comes entirely from how each pipeline structures the model's evidence.}


\textbf{Runtime Performance.} We evaluate runtime cost on selected packages averaging 47.4 files and 2,961.8 lines of code. GuardDog processes packages fastest at 2.55 seconds due to efficient Semgrep execution. OSSGadget requires 7.94 seconds for 133 pattern rules. SAP\_XGB needs 9.58 seconds for feature extraction and inference. Packj\_trace demands 17.90 seconds for dynamic analysis. SocketAI is slowest at 146.34 seconds due to multiple LLM queries. For large-scale scanning, GuardDog processes 1,412 packages per hour while SocketAI processes only 24.

\wbadd{\textit{Ranking-based detection.} EMPHunter clusters the installation 
scripts of newly uploaded packages and surfaces outliers as candidate malware, 
so we evaluate it as a ranking task. Each ranking instance mixes one malicious 
version with 111 benign packages, and we record the rank at which the malicious 
version appears. Reading installation scripts only, the tool covers the 5,236 
malicious versions that declare an install hook (81.6\% of the corpus) and 
leaves the remaining 1,184 outside its design. Within this scope it recalls 
80.46\% at Top-10 with an 8.28\% false-positive rate (mAP@10 of 62.35\%, 
Table~\ref{tab:emphunter_perform}). Top-K precision is capped by construction. 
With one malicious version per instance, at most one of the K inspected 
candidates can be a true positive, so precision cannot exceed $1/K$, and the 
8.05\% at Top-10 sits close to that 10\% ceiling. EMPHunter thus examines the 
spot that code-analysis tools ignore, the install commands inside 
\texttt{package.json}, which makes it a complement to those tools rather than 
a standalone detector.}

\findingbox{\wbadd{Finding 1: Malicious and benign packages invoke the same OS APIs, so every detector must separate \textit{capability} (what code can do) from \textit{intent} (what code aims to do). Tools that flag every sensitive API catch most malware but drown in false positives, while tools that demand
strict evidence stay precise but miss attacks. GuardDog strikes the best balance among conventional tools (93.32\% F1) through end-to-end taint analysis, and IntelGuard goes further (95.98\% F1) by grounding LLM reasoning in examples retrieved from its knowledge base.}}



\subsection{RQ2: Fine-grained Behavioral Analysis}

While RQ1 evaluated overall detection effectiveness, RQ2 examines detection capabilities across specific malicious behaviors and evasion techniques.

\subsubsection{RQ2.1: Behavioral Detection Capability}

\begin{figure}[t]
    \centering
    \includegraphics[width=0.83\linewidth]{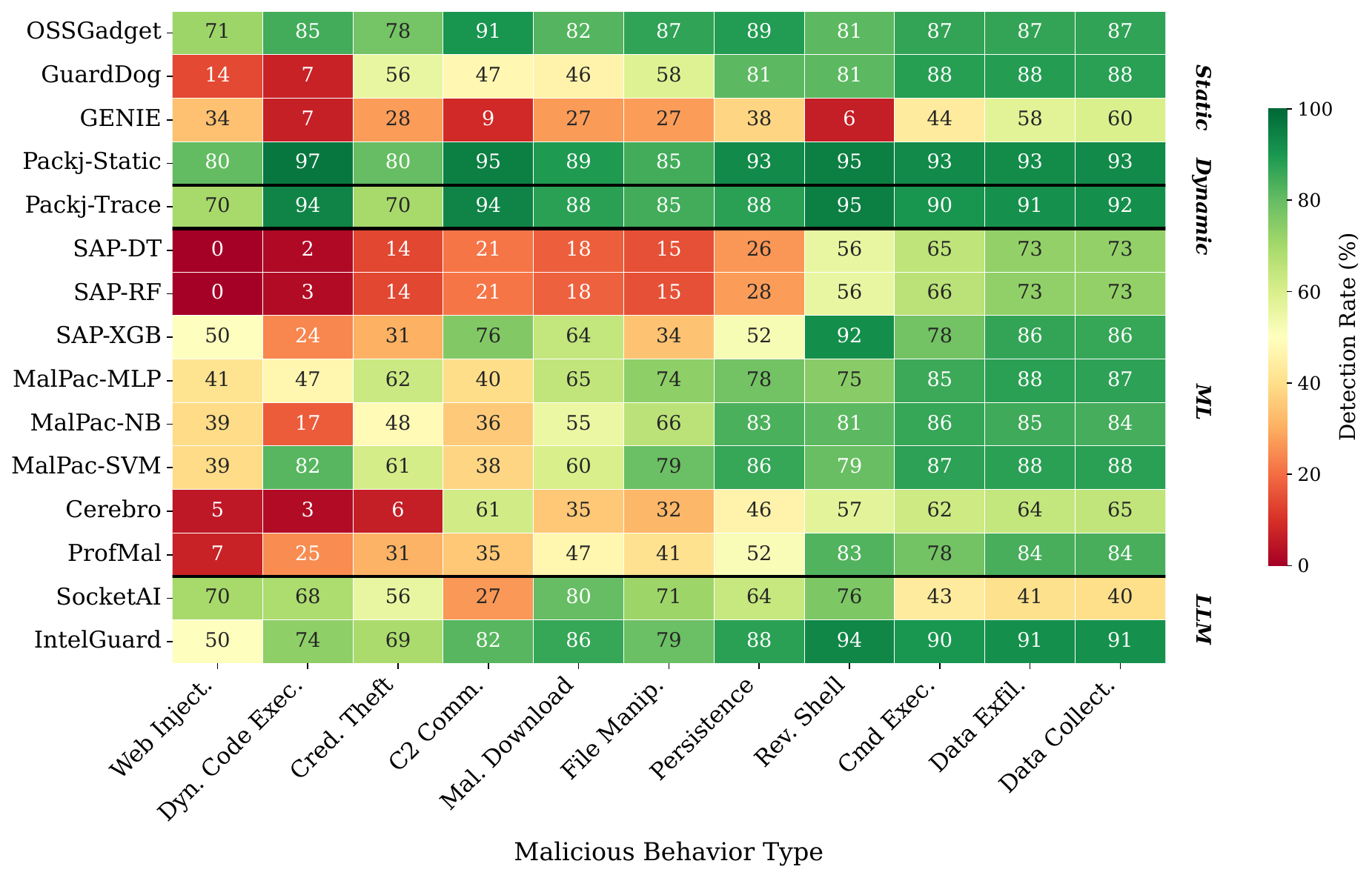}
    \caption{\wbadd{Detection Rates Across Malicious Behaviors}}
    \label{fig:behavior_detection_heatmap}
\end{figure}

\textbf{Results \& Analysis.} Detection performance varies significantly 
across the 11 behavior categories and \wbadd{15} tool variants, as shown in 
Figure~\ref{fig:behavior_detection_heatmap}. Packj\_static achieves the 
broadest coverage, exceeding 80\% for 10 of 11 categories through 
comprehensive API monitoring. In contrast, GENIE\wbadd{,} SAP\_DT/RF\wbadd{, and ProfMal} show 
critical gaps in dynamic code execution, C2 communication, and web injection, 
reflecting the limitations discussed in RQ1. GuardDog exhibits the most uneven profile among 
well-performing tools: it exceeds 94\% for data exfiltration and command 
execution but drops to only 7\% for dynamic code execution and 27\% for 
web injection, because its Semgrep rules were not designed to cover these categories.


\textit{High-detectability behaviors.} Data collection, data exfiltration, 
and command execution achieve mean detection rates above 75\% across all 
tools. Beyond their prevalence, these behaviors benefit from a coupling 
effect: they rarely occur in isolation, and co-occurrence amplifies detection signals. 
When data collection and exfiltration co-occur, mean detection rises from \wbadd{40.0\%} and \wbadd{72.2\%} individually to \wbadd{85.4\%}. 
The effect is most pronounced for SAP\_DT, which jumps from 3.2\% to 79.3\%, 
because a single \textit{os.hostname()} call is ambiguous, but the chain 
\textit{os.hostname()} $\to$ \textit{JSON.stringify()} $\to$ 
\textit{https.request()} constitutes unambiguous intent evidence.

\textit{Low-detectability behaviors.} Dynamic code execution, web injection, 
and credential theft show the widest cross-tool variation and the lowest mean 
detection rates. These behaviors do not form behavioral chains, so there is 
no coupling effect to amplify the signal. More critically, web injection 
targets browser-side DOM manipulation, but all evaluated tools analyze 
server-side Node.js code, creating a structural attack surface mismatch that 
no detection rule can bridge. \wbadd{Even IntelGuard, the strongest detector overall, drops from 98.5\% on data exfiltration to
62.5\% on web injection, which places the limit on the analysis surface rather than on the
sophistication of the reasoning applied to the code.} Credential theft faces a similar problem: the 
APIs involved are identical to those used by legitimate authentication 
libraries, leaving feature-based tools with no discriminative signal.

\findingbox{Finding 2: Detection difficulty is shaped by two structural 
factors. Behavioral coupling amplifies intent signals when behaviors 
co-occur: the collection+exfiltration chain raises SAP\_DT detection from 
3.2\% to 79.3\%. Attack surface misalignment makes certain behaviors 
structurally undetectable: web injection targets browser-side DOM while 
tools analyze Node.js code, and credential theft uses APIs identical to 
legitimate authentication libraries.}

\subsubsection{RQ2.2: Evasion Technique Analysis}

\wbadd{To measure how well each detector copes with evasion, we first label 
which technique each malicious package uses. We adopt eight evasion categories 
established in prior studies of malicious 
packages~\cite{ohm2020backstabber, ladisa2023sok}, from obfuscation to trace 
cleanup. For each category we assemble a set of pattern rules, adapting 
signatures from prior work and public tools and adding our own for the cases 
they miss, 53 rules in total, and run them over each package's malicious code 
and \texttt{package.json} scripts. An expert then reviews every match, and a 
package is labeled evasion-free only when none survives review.}

\begin{figure}[t]
    \centering
    \begin{subfigure}[b]{0.48\linewidth}
        \centering
        \includegraphics[width=\linewidth]{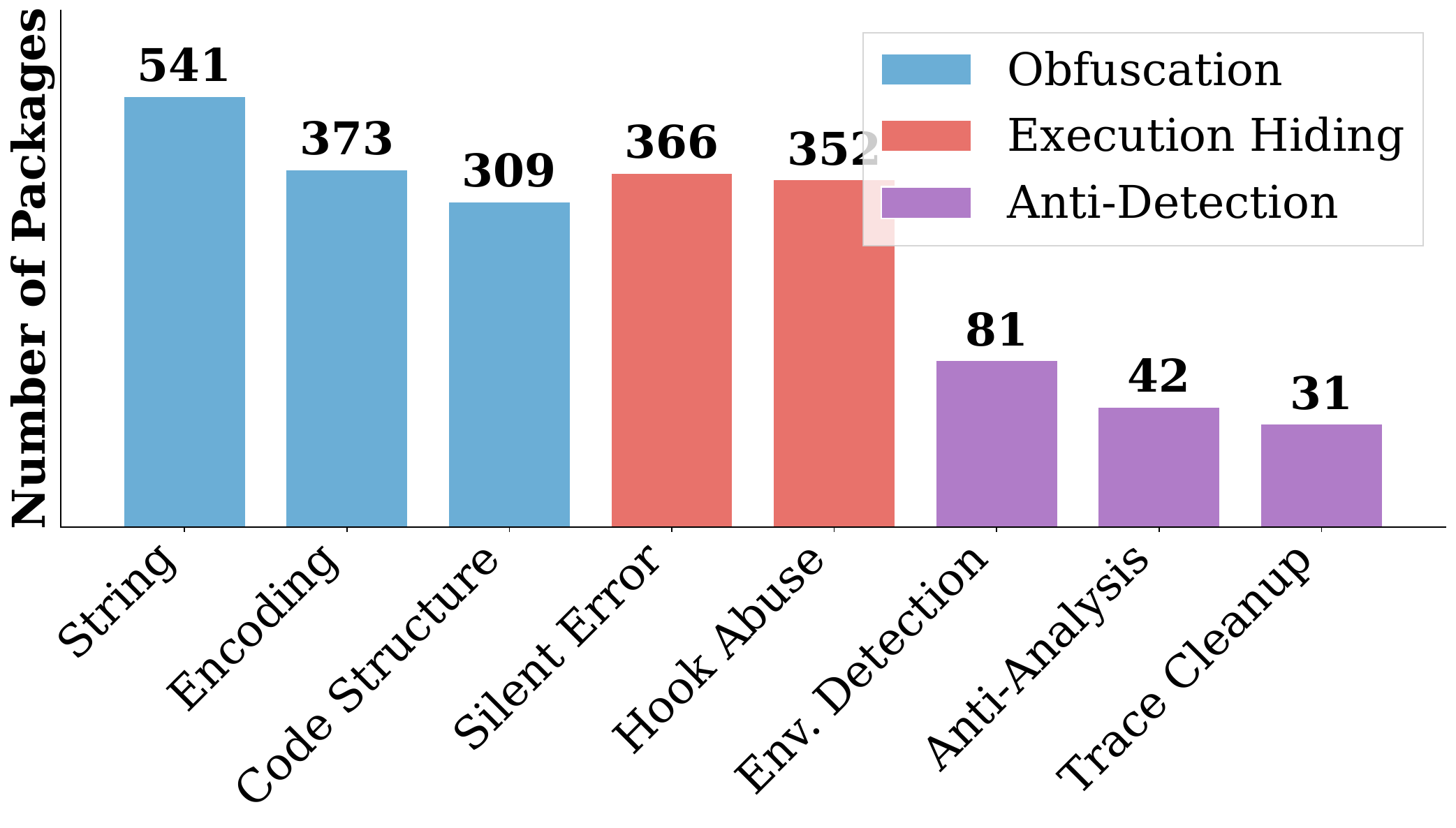}
        \caption{Evasion Distribution}
        \label{fig:evasion_distribution}
    \end{subfigure}
    \hfill
    \begin{subfigure}[b]{0.48\linewidth}
        \centering
        \includegraphics[width=\linewidth]{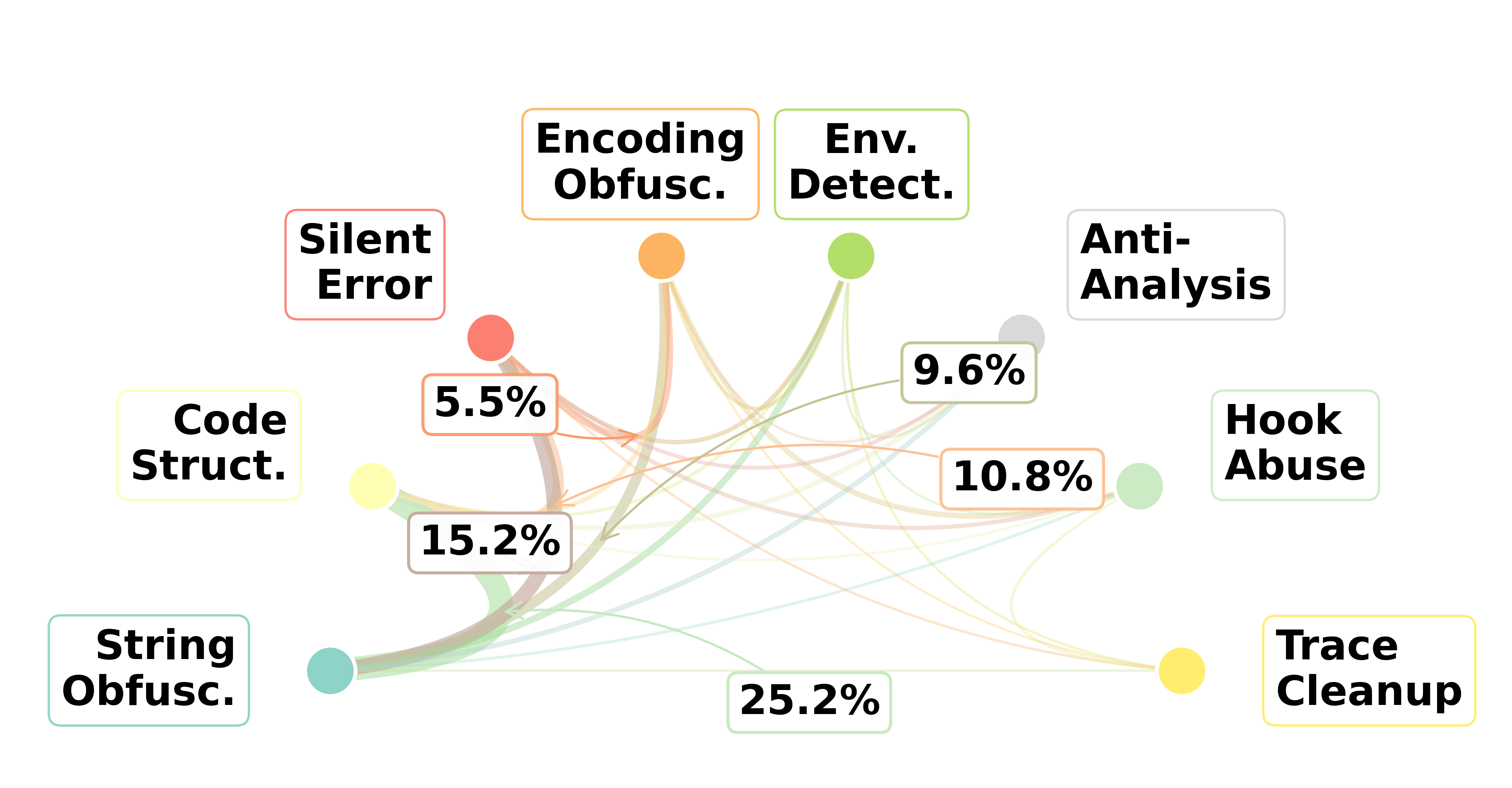}
        \caption{Evasion Co-occurrence}
        \label{fig:evasion_combination}
    \end{subfigure}
    \vspace{5pt}
    \caption{Evasion Technique Distribution}
    \label{fig:evasion_distribution_analysis}
\end{figure}

\begin{figure}[t]
    \centering
    \includegraphics[width=0.83\linewidth]{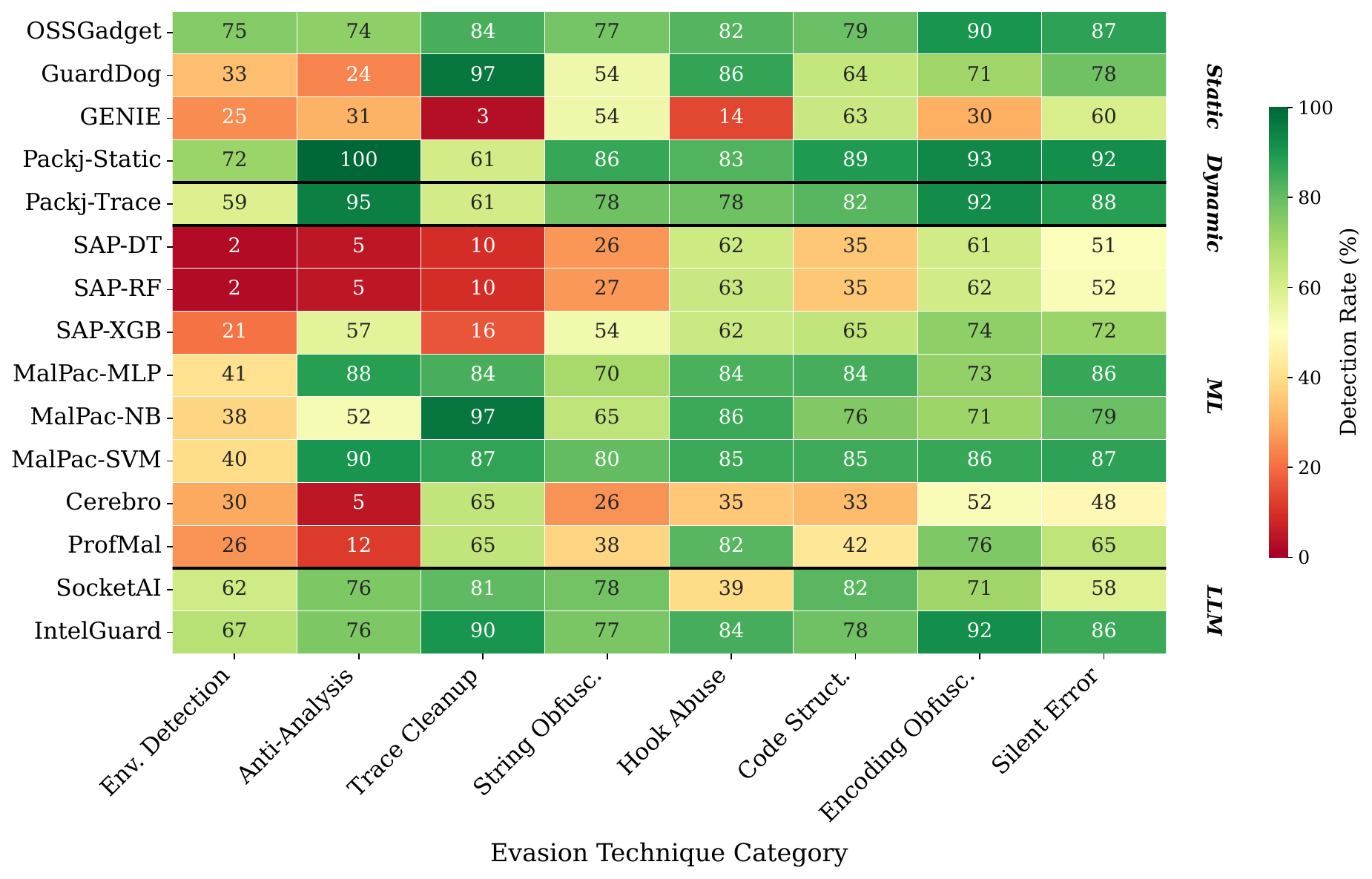}
    \caption{\wbadd{Detection Rates Against Evasion Techniques}}
    \label{fig:evasion_detection_heatmap}
\end{figure}

Figure~\ref{fig:evasion_distribution} shows the distribution of 8 evasion technique categories, and Figure~\ref{fig:evasion_detection_heatmap} presents detection rates against each technique.

\textbf{Results \& Analysis.} Among 6,420 malicious packages, 80.3\% employ 
no evasion technique at all. Among the 19.7\% that do, string obfuscation 
is most prevalent (541 packages), followed by encoding obfuscation, silent 
error handling, and hook abuse. Despite their rarity, environment detection 
and anti-analysis are the hardest to detect, with mean detection rates of 
\wbadd{39.5\%} and \wbadd{52.7\%} respectively.

\textit{Obfuscation techniques.} String and code structure obfuscation defeat 
feature-level tools by altering the surface form from which features are 
computed. Attackers rename identifiers with hex-encoded strings such as 
\textit{\_0x27b98b} and flatten control flow, stripping the statistical 
signals that SAP\_DT/RF and Cerebro rely on, all three falling to 
\wbadd{26.6\%} or below on string obfuscation. Packj\_static holds at 
\wbadd{86.0\%} because it monitors API calls regardless of how identifiers 
are named.


\textit{Anti-analysis techniques.} Anti-analysis exhibits the largest 
detection gap of all categories, from 5\% for SAP\_DT/RF and Cerebro to 
100\% for Packj\_static. Attackers use self-modifying \textit{eval} wrappers 
with integrity checks: if a scanner modifies the payload, the hash changes 
and the code refuses to execute, defeating any tool that instruments or 
rewrites code during analysis. Packj\_static bypasses this entirely by 
monitoring system calls at the OS level, never touching the code itself. 
Trace cleanup shows an equally sharp split: \wbadd{SAP\_DT/RF} and GENIE 
detect below \wbadd{10\%} because evidence is deleted before feature 
extraction completes, while GuardDog and MalPac-NB both reach 97\% through 
pattern matching on file deletion APIs.


\textit{Paradigm-level pattern.} Evasion effectiveness depends on whether 
the technique targets the same abstraction level as the detector. Obfuscation 
defeats feature-level tools but leaves API-level tools unaffected. 
Anti-analysis and environment detection defeat code-level tools by using 
APIs identical to legitimate system queries, but fail against syscall-level 
tools that observe actual OS interactions rather than inspecting code. 
\wbadd{A richer representation does not exempt a tool from this rule. ProfMal 
builds behavior graphs from static and dynamic analysis, yet falls to 11.9\% 
on anti-analysis and 25.9\% on environment detection, tracking SAP and 
Cerebro rather than the syscall-level tools, because those graphs still 
derive from the code surface these techniques alter. What matters is where 
a tool's decision boundary is written. IntelGuard never falls below 66.7\% 
on any technique and averages 81.3\%, within a tenth of a point of 
OSSGadget's 81.2\%. A regular expression and a retrieved precedent both live 
outside the package, so rewriting its surface moves neither.} No single 
technique defeats all abstraction levels.

\textit{Evasion combinations.} Attackers frequently layer techniques, with 
code structure combined with string obfuscation accounting for 36.81\% of 
all combinations. Packages using 2--3 techniques are detected by fewer tools 
on average than those using only one, but packages using 4 or more are 
detected by more tools than any other group, because heavy obfuscation 
produces its own anomalies: elevated entropy, abnormal code structure, and 
large decoding scaffolds. Evasion code intended to hide intent ends up 
signaling that something is being hidden.



\findingbox{Finding 3: Most attackers do not attempt evasion at all, as npm 
performs no scanning before publication and simple attacks remain 
effective. Among those that do, adding techniques offers little protection. 
Packages combining four or more are caught by more tools than those using 
two or three, since heavy obfuscation leaves traces of its own. Attackers 
respond by abandoning failed techniques rather than refining them, moving 
instead to layers that tools do not inspect.}

\subsubsection{RQ2.3: Installation Script Abuse}

\textbf{Results \& Analysis.} Among 6,420 malicious packages, 4,636 
(72.21\%) exploit installation scripts as attack vectors\wbadd{, counting a package only when the 
code confirmed malicious in its \texttt{package.json} is itself a declared hook, so a benign build 
tool that merely declares one never enters the count}. Preinstall scripts 
dominate, appearing in 61.15\% of all malicious packages, followed by 
postinstall (6.74\%) and install scripts (2.20\%). Malicious code 
concentrates in two locations: \textit{index.js} in 46.9\% 
of packages and the \textit{package.json} scripts section in 38.2\%. More 
strikingly, 1,362 packages (21.2\%) rely entirely on install scripts with 
no malicious JavaScript code in any file.

Preinstall scripts are the preferred entry point for a structural reason: 
they execute automatically during \texttt{npm install} with full user 
privileges before dependency resolution, before security tooling can 
activate, and regardless of whether the installation ultimately succeeds. 
The silent execution pattern \texttt{"preinstall": "node index.js > 
/dev/null 2>\&1"} further ensures no console output reveals the attack. 
This is not a vulnerability being exploited but a design decision: npm's 
lifecycle hook mechanism was built on the assumption that publishers are 
benign, while its open publication model provides no such guarantee.

The fundamental detection difficulty is not technical but semantic. 
Legitimate build tools such as \textit{husky}, \textit{patch-package}, 
and \textit{prisma generate} use the exact same preinstall and postinstall 
hooks for entirely benign purposes. A tool that flags all install scripts 
produces massive false positives as Packj does with 6,243 FP; a tool that 
requires stronger evidence of malicious intent misses most attacks as GENIE 
does with only 14.77\% hook abuse detection. The 21.2\% of install-only 
malware sharpens this further: when the entire attack is a single command 
string in \texttt{package.json} invoking a separate JavaScript file, there 
is no obfuscation or suspicious API to detect, just a pattern 
indistinguishable from any legitimate build script.

\findingbox{Finding 4: 72.21\% of malicious packages exploit install 
scripts because npm lifecycle hooks guarantee execution without sandboxing. 
The 21.2\% that embed attacks entirely in \texttt{package.json} expose an 
architectural blind spot: every code-analysis tool assumes malicious 
behavior lives in code, and when it does not, no rule refinement helps.}






\subsection{RQ3: Temporal Evolution Analysis}

\begin{figure}[t]
    \centering
    \includegraphics[width=0.3\textwidth]{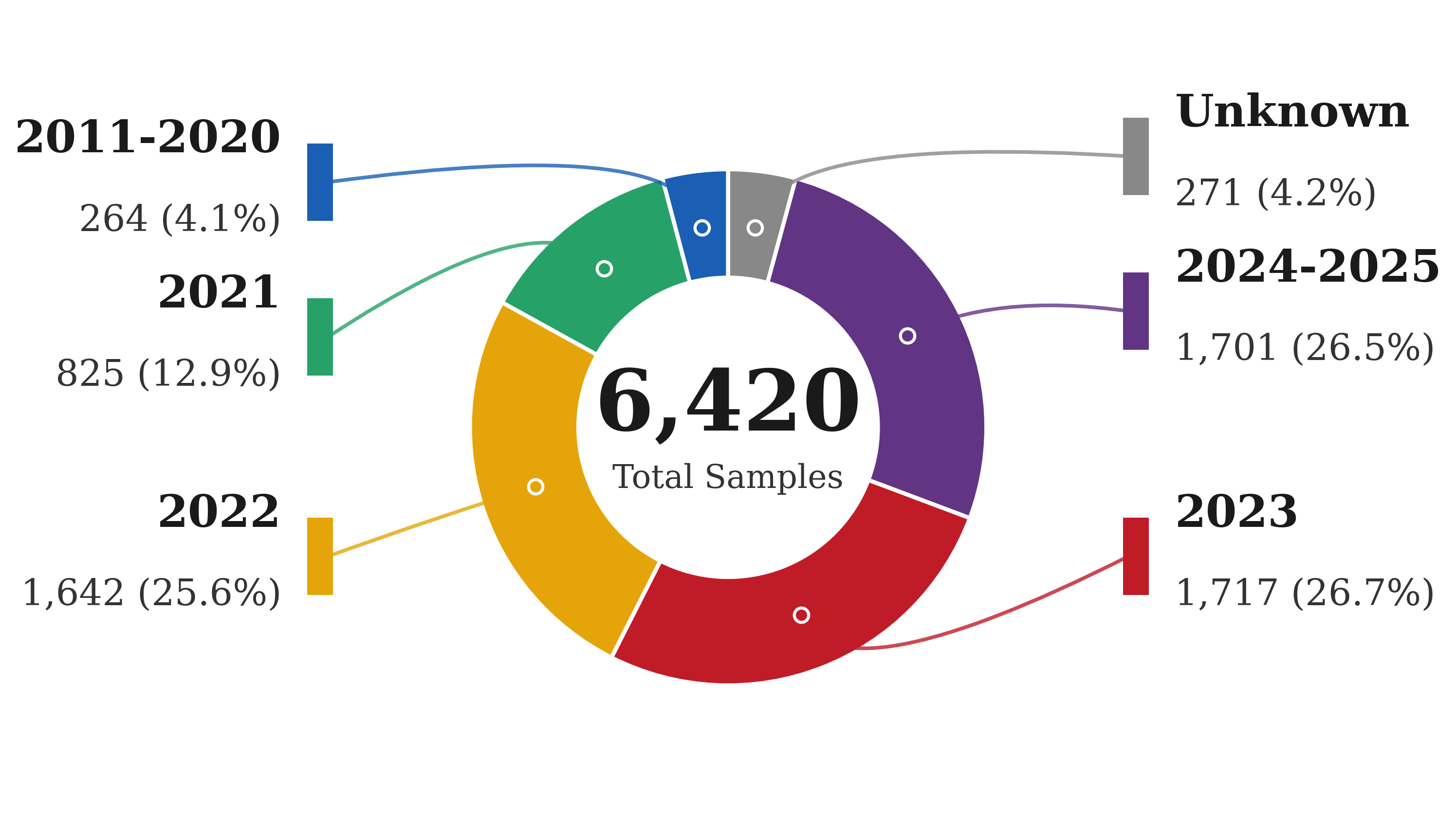}
    \caption{Package Temporal Distribution}
    \label{fig:temporal_distribution}
    \vspace{5pt}
\end{figure}


\begin{table}[t]
    \centering
    \caption{Evasion Techniques Across Time Periods}
    \label{tab:evasion_evolution}
    \scriptsize
    \setlength{\tabcolsep}{4pt}
    \renewcommand{\arraystretch}{1.0}
    \resizebox{\columnwidth}{!}{%
    \begin{tabular}{@{}lccccc|c@{}}
    \toprule
    \multirow{2}{*}{\textbf{Evasion Technique}} & \multicolumn{5}{c|}{\textbf{Time Period}} & \multirow{2}{*}{\textbf{Total}} \\
    \cmidrule(lr){2-6}
     & \textbf{$\leq$2020 (249)} & \textbf{2021 (817)} & \textbf{2022 (1,596)} & \textbf{2023 (1,704)} & \textbf{2024-25 (1,622)} & \\
    \midrule
    String Obfuscation & 31 (12.4\%) & 77 (9.4\%) & 149 (9.3\%) & 104 (6.1\%) & 150 (9.2\%) & 541 \\
    Encoding Obfuscation & 48 (19.3\%) & 136 (16.6\%) & 47 (2.9\%) & 90 (5.3\%) & 37 (2.3\%) & 373 \\
    Code Structure Obfusc. & 10 (4.0\%) & 63 (7.7\%) & 102 (6.4\%) & -- & 102 (6.3\%) & 309 \\
    Silent Error Handling & -- & 59 (7.2\%) & 112 (7.0\%) & 63 (3.7\%) & 106 (6.5\%) & 366 \\
    Hook Abuse & 12 (4.8\%) & 55 (6.7\%) & 59 (3.7\%) & 49 (2.9\%) & 152 (9.4\%) & 352 \\
    Environment Detection & 29 (11.6\%) & -- & -- & -- & -- & 81 \\
    Anti-Analysis & -- & -- & -- & -- & -- & 42 \\
    Trace Cleanup & -- & -- & -- & 21 (1.2\%) & -- & 31 \\
    \bottomrule
    \end{tabular}%
    }
    \scriptsize
    \textit{Note: Only top-5 evasion techniques per period are shown. ``--'' indicates the technique was not among the top five for that period.}
\end{table}

While RQ1 and RQ2 evaluated detection effectiveness on our complete dataset, RQ3 examines whether detection tools maintain consistent performance as attack techniques evolve over time.

\textbf{Timestamp Collection.} We collect publication timestamps via 
the NPM Registry API, retrieving timestamps for 6,149 of 6,420 packages 
(98.5\%). As shown in Figure~\ref{fig:temporal_distribution}, 95.7\% 
of packages were published between 2021 and 2025, reflecting the 
growing threat of supply chain attacks and increased detection efforts 
in recent years.



\subsubsection{Detection Performance Evolution}

\begin{figure}
    \centering
    \includegraphics[width=1.0\linewidth]{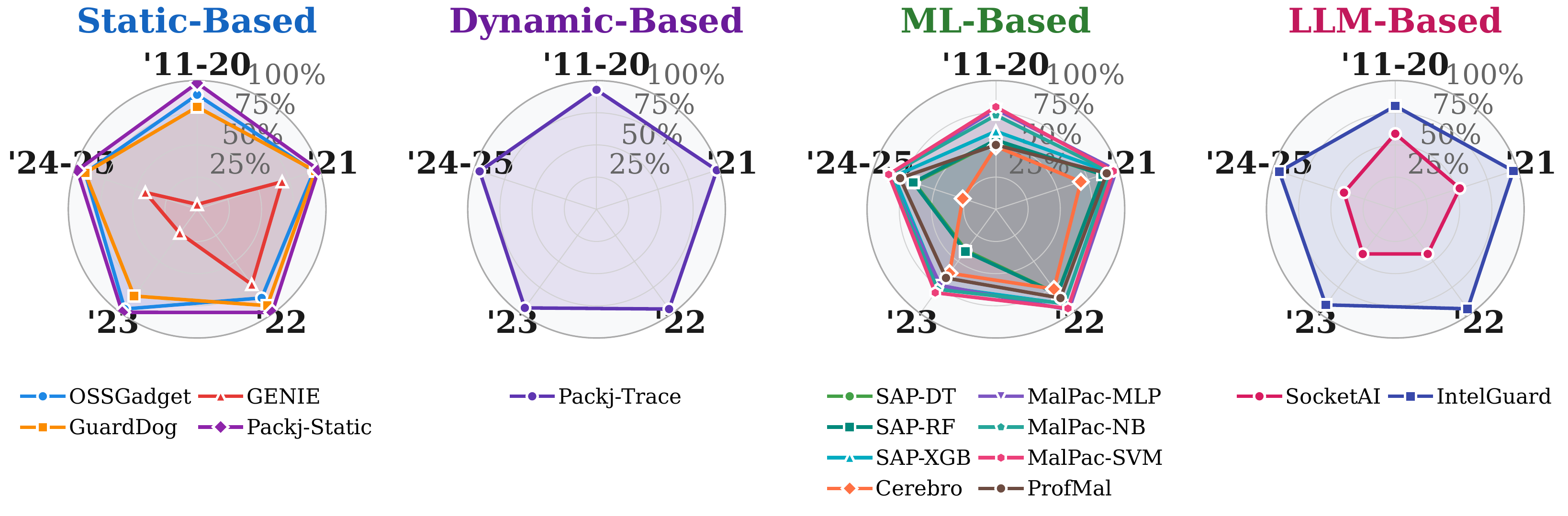}
    \caption{\wbadd{Detection Rate Comparison Across Time Periods by Tool Category}}
    \label{fig:detection_rate_overtime}
\end{figure}

Figure~\ref{fig:detection_rate_overtime} presents detection rate trends across time periods for all \wbadd{15} tool variants.

\textbf{Results \& Analysis.} Detection tools exhibit divergent temporal 
patterns. Packj\_static maintains above 97\% across all periods with only 
$\pm$1.15\% variance\wbadd{, and IntelGuard stays above 91\% from 2021 onward}, while SAP\_DT drops from 87.15\% in 2021 to 39.49\% 
in 2023, \wbadd{ProfMal falls from 90.42\% to 65.93\% over the same years,} Cerebro collapses from 76.61\% to 27.04\% between 2022 and 
2024--2025, and GENIE oscillates between 3.79\% and 72.17\%. All tools 
use fixed detection parameters, so the divergence reflects changes in the 
malware population, not tool updates.

The ML degradation contradicts the conventional concept drift explanation. 
Between 2022 and 2023, obfuscation decreased and malware became simpler, 
not more sophisticated. SAP\_DT collapsed because minimal-footprint malware 
became statistically indistinguishable from benign packages in feature 
space, a phenomenon we term \textit{concept convergence}. \wbadd{ProfMal, 
published four years later and trained on a more recent corpus, traces the same 
trough from 90.42\% to 65.93\%, which shows that convergence is a property of the 
feature space rather than a defect of any one model.} Its partial 
recovery in 2024--2025 was not adaptation: the hook-abuse surge happened 
to trigger SAP's install-script feature, shifting the malware population 
back into its coverage. Cerebro's collapse follows a different mechanism: 
its BERT embeddings peak on 2022 malware because its training corpus aligns 
with that era, then fail silently as attack code evolves beyond its learned 
representations. GENIE's 68-point oscillation reflects a third failure mode: 
not degradation but query-pattern misalignment, peaking when dominant 
attacks match its CodeQL conditions and collapsing when they do not. 
GuardDog is the only tool showing genuine recovery, demonstrating that 
pattern-based tools can adapt through rule maintenance in a way that 
ML models fundamentally cannot. \wbadd{IntelGuard remains stable without any rule 
maintenance because it grounds each judgment in retrieved precedent rather than in 
learned surface statistics, so shifts in code appearance leave its evidence base 
intact.} Underlying all these patterns is a single 
principle: tools that detect what code \textit{does} are temporally robust 
because attack objectives are invariant, while tools that detect code 
surface properties are fragile because appearances evolve freely.

\subsubsection{Evasion Technique Evolution}

Table~\ref{tab:evasion_evolution} presents the evolution of evasion techniques across 8 categories from 2011 to 2025.



\textbf{Results.} Evasion use grows from 94 packages in 2011--2020 to 358 in 
2024--2025, a 3.8x increase, but the corpus grows 6.5x over the same span (249 
to 1,622), so the share of malware carrying any evasion falls from 37.8\% to 
22.1\%. Encoding obfuscation dominated early malware but declined from 19.3\% 
to 2.3\%. String obfuscation stays between 6.1\% and 12.4\% across all periods. 
Hook abuse rebounds from 2.9\% in 2023 to 9.4\% in 2024--2025, the most 
prevalent technique in that period. Anti-analysis stays rare throughout.

\textbf{Analysis.} Encoding obfuscation gives way to hook abuse across the full 
span, 19.3\% $\rightarrow$ 2.3\% for the former and 4.8\% $\rightarrow$ 9.4\% 
for the latter. Hook abuse exploits npm's trusted execution model, where 
\textit{preinstall}/\textit{postinstall} scripts run with full privileges, 
bypassing code-level analysis entirely. The decline of encoding obfuscation 
explains the ML collapse. SAP models trained on 2021 data learned high-entropy 
patterns as malicious indicators, and once attackers moved to clean code with 
standard APIs, those features stopped separating malicious from benign, causing 
the 47.66 percentage point drop in 2023. \wbadd{ProfMal follows the same path 
despite a more recent training corpus, so the shift defeats the feature space 
rather than one model.} GuardDog recovers to 91.42\% because its Semgrep rules 
match manifest-level hook patterns that other tools miss\wbadd{, and IntelGuard 
holds above 91\% because retrieved precedent does not age as code appearance 
changes}.

\findingbox{Finding 5: \wbadd{Temporal robustness follows the origin of a detector's decision 
boundary, not its paradigm. Every boundary fitted to a corpus ages, and ProfMal proves that 
representation richness is no defense, since its behavior graph troughs as steeply as SAP's 
statistics, while IntelGuard stays above 91\% by grounding verdicts in retrieved evidence. SAP and Cerebro age by different routes, through convergence toward benign code and through stale embeddings, and GuardDog alone recovered, through rule updates rather than retraining.}}


\subsubsection{Sensitive API Evolution}

\begin{figure}
    \centering
    \includegraphics[width=1.0\linewidth]{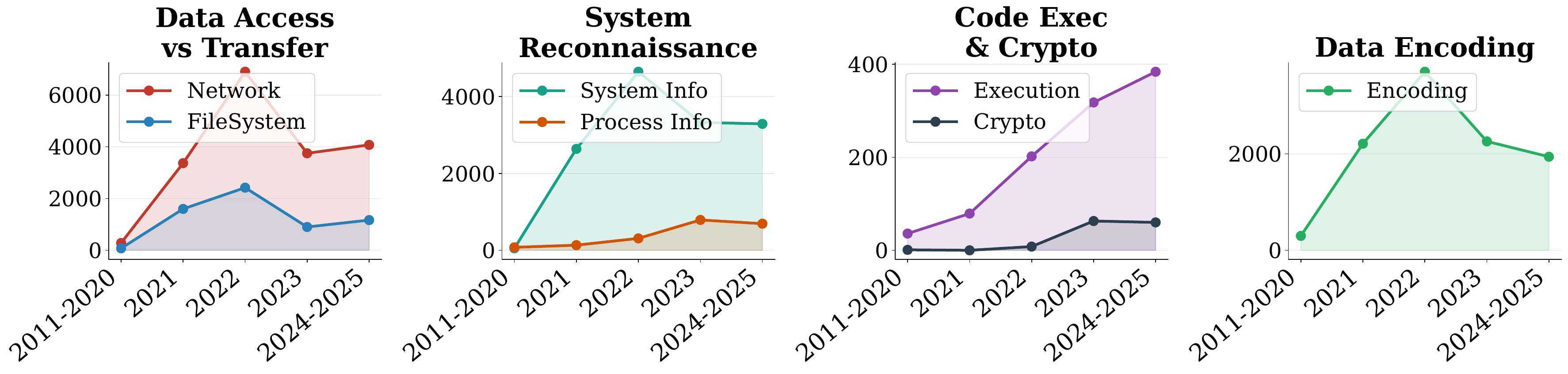}
    \caption{Evolution of Sensitive API Usage in Malicious NPM Packages Over Time}
    \label{fig:api_trends}
\end{figure}

To understand how malicious code patterns evolve beyond evasion techniques, we analyze the sensitive APIs invoked by malicious packages across time periods.

\textbf{API Extraction.} We extract sensitive API calls from malicious code snippets using \textit{tree-sitter}~\cite{tree_sitter}, a parsing library that constructs abstract syntax trees (ASTs) for JavaScript code. For each code snippet, we parse the AST and identify API invocations through three mechanisms: (1) call expressions matching known sensitive functions, (2) member expressions accessing module methods (e.g., \textit{fs.readFile}, \textit{os.hostname}), and (3) string literals containing API names to capture obfuscated dynamic calls. Following the sensitive API taxonomy established in prior work~\cite{duan2020towards, zhang2025killing}, we categorize extracted APIs into seven categories: network, filesystem, system\_info, process\_info, execution, encoding, and crypto. This extraction successfully processes 6,239 packages, identifying 53,576 API invocations.

\textbf{Results \& Analysis.} Figure~\ref{fig:api_trends} presents the 
temporal evolution of sensitive API usage across 6,239 packages. Most 
categories peaked in 2022 and declined thereafter, while execution APIs 
grew continuously across all periods. The average API count per package 
declined from 12.52 in 2021 to 6.82 in 2023, and by 2024--2025, 12.6\% 
of malware invokes only 1--2 APIs total, up from 5.1\% in 2021, with 51\% 
of these minimal packages importing only \textit{child\_process} to execute 
a single shell command. Crypto APIs emerged after 2022 and have grown 
steadily, suggesting increasing use of encrypted communication channels. 
This reflects a structural shift from 2021-era full reconnaissance attacks, 
which collected hostname, environment variables, and user information before 
exfiltrating via \textit{https.request}, toward single-step execution: when 
the network operation is embedded inside a shell command such as 
\texttt{exec(`curl attacker.com | sh`)}, API-level tools tracking 
\textit{https.request} as a network sink cannot observe it at all, and the 
behavioral coupling effect that amplifies detection signals disappears 
entirely.

\findingbox{Finding 6: Sensitive API usage has shifted from broad 
multi-category reconnaissance toward minimal execution-focused attacks. 
Single-step shell command attacks eliminate the behavioral coupling that 
amplifies detection signals and bypass API-level sink definitions entirely, 
leaving process-execution-level monitoring as the only reliable detection 
surface against this pattern.}

\subsection{RQ4: Tool Complementarity Analysis}






While RQ1-RQ3 revealed significant performance variations among individual tools, RQ4 examines whether strategic tool combinations can overcome these limitations. We evaluate pairwise combinations using two strategies: union (flagging packages detected by either tool) and intersection (requiring both tools to agree).

\begin{figure}
    \centering
    \includegraphics[width=1\linewidth]{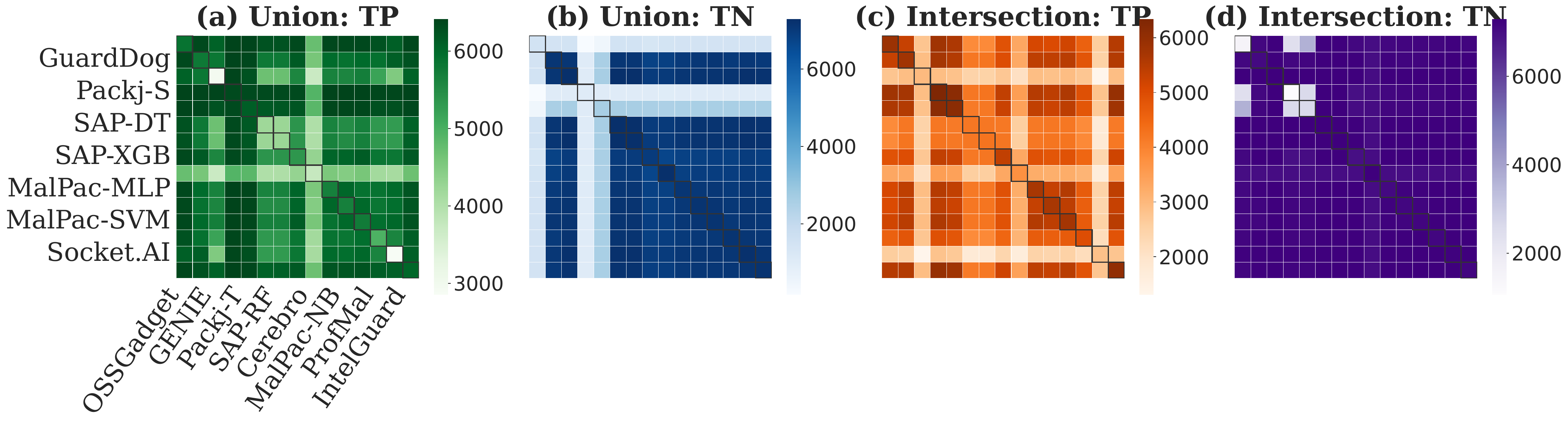}
    \caption{\wbadd{Complementary Analysis of Detection Tools: TP/TN Performance Matrix}}
    \label{fig:tool_combination_heatmap}
    \vspace{5pt}
\end{figure}

\begin{table}[t]
    \centering
    \caption{\wbadd{Top Tool Combinations and Their Gain over the Stronger Single Tool}}
    \label{tab:tool_combinations}
    \small
    \resizebox{\columnwidth}{!}{%
    \begin{tabular}{@{}lcccc|lcccc@{}}
    \toprule
    \multicolumn{5}{c|}{\textbf{Union Strategy}} & \multicolumn{5}{c}{\textbf{Intersection Strategy}} \\
    \cmidrule(r){1-5} \cmidrule(l){6-10}
    \textbf{Combination} & \textbf{Prec.} & \textbf{Rec.} & \textbf{F1} & \textbf{\wbadd{$\Delta$}} & \textbf{Combination} & \textbf{Prec.} & \textbf{Rec.} & \textbf{F1} & \textbf{\wbadd{$\Delta$}} \\
    \midrule
    \wbadd{IntelGuard + MalPac\_SVM} & \wbadd{96.81} & \wbadd{97.24} & \wbadd{\textbf{97.02}} & \wbadd{+1.05} & \wbadd{Packj\_static + IntelGuard} & \wbadd{98.48} & \wbadd{92.66} & \wbadd{\textbf{95.48}} & \wbadd{$-$0.49} \\
    \wbadd{GuardDog + IntelGuard} & \wbadd{96.04} & \wbadd{97.41} & \wbadd{96.72} & \wbadd{+0.74} & \wbadd{Packj\_trace + IntelGuard} & \wbadd{98.72} & \wbadd{90.23} & \wbadd{94.29} & \wbadd{$-$1.69} \\
    \wbadd{IntelGuard + MalPac\_NB} & \wbadd{96.65} & \wbadd{96.74} & \wbadd{96.70} & \wbadd{+0.72} & \wbadd{GuardDog + Packj\_static} & \wbadd{97.39} & \wbadd{89.07} & \wbadd{93.04} & \wbadd{$-$0.28} \\
    \wbadd{SAP\_RF + IntelGuard} & \wbadd{98.21} & \wbadd{94.89} & \wbadd{96.52} & \wbadd{+0.55} & \wbadd{GuardDog + IntelGuard} & \wbadd{99.48} & \wbadd{86.32} & \wbadd{92.44} & \wbadd{$-$3.54} \\
    \midrule
    \wbadd{GuardDog + SocketAI} & \wbadd{96.45} & \wbadd{95.14} & \wbadd{95.79} & \wbadd{+2.47} & \wbadd{Packj\_static + MalPac\_SVM} & \wbadd{98.23} & \wbadd{87.27} & \wbadd{92.43} & \wbadd{+8.23} \\
    \bottomrule
    \end{tabular}%
    }
    \footnotesize
    \wbadd{\textit{Note: $\Delta$ is the F1 gain over the stronger of the two tools used alone. Union flags a package when either tool detects it; Intersection requires both. In each column, the first four rows are the top pairs by F1, and the row below the rule is the strongest IntelGuard-free pair not among them, shown for reference.}}
\end{table}

\textbf{Results \& Analysis.} Table~\ref{tab:tool_combinations} and 
Figure~\ref{fig:tool_combination_heatmap} present the top combinations 
across all \wbadd{15} tool variants. For union, \wbadd{IntelGuard + MalPac\_SVM achieves the 
best F1 at 97.02\% with 97.24\% recall}, above any individual tool. 
For intersection, \wbadd{Packj\_static + IntelGuard} leads with \wbadd{95.48\%} F1 and 
\wbadd{98.48\%} precision. Cross-methodology pairs consistently outperform 
same-methodology ones, yet paradigm diversity is a proxy for complementarity 
rather than its cause.

Two opposing forces govern combination effectiveness, detection complementarity 
and false-positive introduction. We quantify complementarity with McNemar's 
$\chi^2$. GuardDog and SocketAI reach $\chi^2 = 1{,}920$, while SAP\_DT and 
SAP\_RF reach only 39, because both extract identical features and disagree on 
almost nothing. High $\chi^2$ is necessary but not sufficient. GuardDog and 
Packj\_static reach $\chi^2 = 4{,}020$ yet fail in union, because Packj\_static 
adds only 665 true positives while introducing 6,015 false positives, collapsing 
combined precision from 97\% to 50\%. What separates a useful partner from a 
useless one is therefore the detection mechanism, not the paradigm label. The 
intra-ML pair Cerebro + MalPac\_SVM succeeds because Cerebro reads BERT 
embeddings while MalPac\_SVM reads 24 binary AST features, so the packages each 
tool misses are largely different. SAP\_DT + SAP\_RF pairs two different 
classifiers over the same 140 features, misses almost identical packages, and 
gains nothing from combination. The same accounting explains a counterintuitive 
result. \wbadd{IntelGuard and GuardDog, the two strongest individual tools, are} 
on average hurt by union at \wbadd{$-$0.05 and $-$0.04 F1}, because most partners 
introduce more false positives than additional true positives, while SocketAI 
benefits most at \wbadd{$+$0.08} F1.

For intersection, \wbadd{Packj\_static + IntelGuard} succeeds where union fails because 
requiring both tools to agree filters out each tool's independent false 
positives, \wbadd{IntelGuard grounds its judgment in retrieved precedent} while Packj\_static flags 
legitimate system calls, but packages flagged by both are almost always 
genuine malware.

\findingbox{Finding 7: \wbadd{IntelGuard + MalPac\_SVM achieves the best union F1 (97.02\%), and 
the gain over the stronger tool alone shrinks as that tool grows stronger, from $+$2.47 for the 
best IntelGuard-free pair down to $+$1.05.} The principle 
is asymmetric: the strongest tool degrades because partners add more false 
positives than new true positives, while the weakest tool benefits most 
because a strong partner fills its blind spots at low cost.}

\section{Implications}
\label{sec:implications}


\textbf{For Practitioners.} \wbadd{Tool selection is a budget decision rather than a single
recommendation. Where an LLM budget exists, IntelGuard + MalPac\_SVM union is the most accurate
configuration we measured at 97.02\% F1, and a tiered deployment keeps its cost bounded, since a
cheap pass such as GuardDog screens every package at 2.55\,s each while the LLM verifies only the
flagged minority. Where no LLM budget exists, Cerebro + MalPac\_SVM union reaches 95.15\% F1
entirely offline with no API dependency. Pipelines that cannot tolerate false positives should
prefer intersection, where GuardDog + Packj\_static reaches 93.04\% F1 at 97.39\% precision using
two static tools.} As hook abuse surged to 9.4\% in 2024--2025, organizations should consider 
disabling script execution via \texttt{npm install --ignore-scripts} for untrusted packages. Teams unwilling to retrain ML-based tools quarterly should switch to behavioral detection tools, which remain stable over time.

\textbf{For Researchers.} The capability-intent gap calls for mechanisms that 
supply partial intent evidence without the strict formal verification that 
limits recall. Taint analysis and behavior-chain detection resolve the 
ambiguity through data-flow evidence rather than surface-level features. 
\wbadd{Retrieved precedent is a third such mechanism, and it is 
knowledge-driven rather than data-driven. IntelGuard pairs malicious code with 
the expert reasoning for why that behavior is malicious, so the model judges 
semantics against a prior intent judgment instead of matching syntax or a 
boundary learned from a corpus, and RQ3 finds such a judgment does not age. 
Reasoning is no substitute for that knowledge, since SocketAI's agentic loop of 
three self-critical rounds costs 57$\times$ GuardDog's latency and still loses 
49 recall points to one grounded call.} Dynamic code execution, web injection, and credential theft remain 
structurally underdetected, the most pressing open problems.


\section{Threats to Validity}
\label{sec:threats}

\textbf{Internal Validity.} Ground truth depends on expert annotation; 
three experts achieved 97.4\% consensus, but edge cases may be mislabeled. 
Tools use default parameters for reproducibility; different settings could 
affect results. \wbadd{SocketAI and Cerebro are not open-source. We 
re-implemented SocketAI from its paper~\cite{zahan2024leveraging}, so its 
results may be lower than Socket's production service, and we evaluated 
Cerebro with help from its authors.} LLM-based extraction achieves 97.2\% accuracy on 500 
validated samples, and the remaining 2.8\% errors may affect behavioral 
analysis.

\textbf{External Validity.} The dataset concentrates on packages from
2021--2025 (95.7\%), for which per-period samples exceed 800 and the temporal findings are
correspondingly robust. \wbadd{Finally, our benign set consists of highly downloaded packages. We chose them because their benign labels are the most reliable, since malware in a widely used package would be discovered quickly, and because they trigger the highest false-positive rates, which makes them the most demanding test of precision. Precision on the long tail of rarely downloaded packages, which we do not cover, may therefore differ.}

\textbf{Construct Validity.} The behavioral taxonomy derives from LLM summaries and clustering. Alternative clustering parameters or different embedding models could produce different category boundaries. Evasion technique categorization involves subjective judgment. Some techniques overlap (e.g., encoding obfuscation and string obfuscation), and boundary definitions may vary across researchers, potentially introducing selection bias.

\section{Conclusion}
\label{sec:conclusion}

This paper evaluates \wbadd{11} NPM detection tools with \wbadd{16} 
variants on a unified dataset of 6,420 malicious and 7,288 benign packages. 
\wbadd{IntelGuard achieves the best F1 at 95.98\%, GuardDog the best balance among 
conventional tools at 93.32\%}, and strategic combinations 
reach up to \wbadd{97.21}\% accuracy. ML-based tools degrade severely 
because malware converged toward benign code as obfuscation became unnecessary 
in an unscanned ecosystem, not because attacks grew more sophisticated. 
Our mechanism analysis traces precision-recall trade-offs, temporal 
fragility, and combination effectiveness to one root cause, the ambiguity 
between code capability and malicious intent. Tools anchored to attack 
objectives stay stable while tools anchored to attack implementations do not, 
so detection should be designed around what malware must do.

\section{Data Availability}

Our dataset and evaluation framework are available online: \url{https://doi.org/10.6084/m9.figshare.31869370}.

\begin{acks}
This research is supported by the National Research Foundation, Singapore, and DSO National Laboratories under the AI Singapore Programme (AISG Award No: AISG4-GC-2023-008-1B); by the National Research Foundation Singapore and the Cyber Security Agency under the National Cybersecurity R\&D Programme (NCRP25-P04-TAICeN); and by the Prime Minister's Office, Singapore under the Campus for Research Excellence and Technological Enterprise (CREATE) Programme.
Any opinions, findings and conclusions, or recommendations expressed in these materials are those of the author(s) and do not reflect the views of the National Research Foundation, Singapore, Cyber Security Agency of Singapore, Singapore.
\end{acks}

\clearpage

\bibliographystyle{ACM-Reference-Format}
\bibliography{ref}

@Misc{npmstats,
  title = {NPMJS},
  howpublished = {\url{https://www.npmjs.com/}},
  note = {Accessed: 2025-07-08},
  year = {2025}
}

@Misc{broadcom_malicious_npm,
  title = {Malicious coa and rc npm Packages Discovered},
  author = {Broadcom},
  howpublished = {\url{https://www.broadcom.com/support/security-center/protection-bulletin/malicious-coa-and-rc-npm-packages-discovered}},
  note = {Accessed: 2025-07-08},
  year = {2021}
}

@Misc{cisa_npm_malware,
  title = {Malware Discovered in Popular npm Package ua-parser-js},
  author = {CISA},
  howpublished = {\url{https://www.cisa.gov/news-events/alerts/2021/10/22/malware-discovered-popular-npm-package-ua-parser-js}},
  note = {Accessed: 2025-07-08}
}

@misc{datadog_guarddog,
  author       = {DataDog},
  title        = {GuardDog: A Threat Detection Tool},
  howpublished = {\url{https://github.com/DataDog/guarddog}},
  year         = {2022},
  note         = {Accessed: 2025-07-08}
}

@misc{datadog_malicious_software_packages,
  author       = {Datadog Security Labs},
  title        = {Open-Source Dataset of Malicious Software Packages},
  howpublished = {\url{https://github.com/DataDog/malicious-software-packages-dataset/tree/main/samples/npm}},
  year         = {2023},
  note         = {Accessed: 2025-07-08}
}

@inproceedings{duan2020towards,
  title={Towards Measuring Supply Chain Attacks on Package Managers for Interpreted Languages},
  author={Duan, Ruian and Alrawi, Omar and Kasturi, Ranjita Pai and Elder, Ryan and Saltaformaggio, Brendan and Lee, Wenke},
  booktitle={Network and Distributed Systems Security (NDSS) Symposium 2021},
  year={2021},
  doi={10.14722/ndss.2021.23055}
}

@inproceedings{halder2024malicious,
author = {Halder, Sajal and Bewong, Michael and Mahboubi, Arash and Jiang, Yinhao and Islam, Md Rafiqul and Islam, Md Zahid and Ip, Ryan HL and Ahmed, Muhammad Ejaz and Ramachandran, Gowri Sankar and Ali Babar, Muhammad},
title = {Malicious Package Detection using Metadata Information},
year = {2024},
isbn = {9798400701719},
publisher = {Association for Computing Machinery},
address = {New York, NY, USA},
url = {https://doi.org/10.1145/3589334.3645543},
doi = {10.1145/3589334.3645543},
booktitle = {Proceedings of the ACM Web Conference 2024},
pages = {1779–1789},
numpages = {11},
location = {Singapore, Singapore},
series = {WWW '24}
}

@inproceedings{huang2024spiderscan,
author = {Huang, Yiheng and Wang, Ruisi and Zheng, Wen and Zhou, Zhuotong and Wu, Susheng and Ke, Shulin and Chen, Bihuan and Gao, Shan and Peng, Xin},
title = {SpiderScan: Practical Detection of Malicious NPM Packages Based on Graph-Based Behavior Modeling and Matching},
year = {2024},
isbn = {9798400712487},
publisher = {Association for Computing Machinery},
address = {New York, NY, USA},
url = {https://doi.org/10.1145/3691620.3695492},
doi = {10.1145/3691620.3695492},
booktitle = {Proceedings of the 39th IEEE/ACM International Conference on Automated Software Engineering},
pages = {1146–1158},
numpages = {13},
location = {Sacramento, CA, USA},
series = {ASE '24}
}

@Misc{medium_event_stream,
  title = {Compromised npm Package event-stream},
  author = {Intrinsic},
  howpublished = {\url{https://medium.com/intrinsic-blog/compromised-npm-package-event-stream-d47d08605502}},
  year = {2018},
  note = {Accessed: 2025-07-08}
}

@Misc{rptu_npm_trojan,
  title = {Remote Access Trojaner in npm Paket},
  author = {RPTU Kaiserslautern-Landau},
  howpublished = {\url{https://rptu.de/en/informationssicherheit/sicherheitswarnungen/details/news/remote-access-trojaner-in-npm-paket}},
  year = {2025},
  note = {Accessed: 2025-07-08}
}

@inproceedings{ladisa2023feasibility,
author = {Ladisa, Piergiorgio and Ponta, Serena Elisa and Ronzoni, Nicola and Martinez, Matias and Barais, Olivier},
title = {On the Feasibility of Cross-Language Detection of Malicious Packages in npm and PyPI},
year = {2023},
isbn = {9798400708862},
publisher = {Association for Computing Machinery},
address = {New York, NY, USA},
url = {https://doi.org/10.1145/3627106.3627138},
doi = {10.1145/3627106.3627138},
booktitle = {Proceedings of the 39th Annual Computer Security Applications Conference},
pages = {71–82},
numpages = {12},
location = {Austin, TX, USA},
series = {ACSAC '23}
}

@inproceedings{li2023malwukong,
author = {Li, Ningke and Wang, Shenao and Feng, Mingxi and Wang, Kailong and Wang, Meizhen and Wang, Haoyu},
title = {MalWuKong: Towards Fast, Accurate, and Multilingual Detection of Malicious Code Poisoning in OSS Supply Chains},
year = {2024},
isbn = {9798350329964},
publisher = {IEEE Press},
url = {https://doi.org/10.1109/ASE56229.2023.00073},
doi = {10.1109/ASE56229.2023.00073},
booktitle = {Proceedings of the 38th IEEE/ACM International Conference on Automated Software Engineering},
pages = {1993–2005},
numpages = {13},
location = {Echternach, Luxembourg},
series = {ASE '23}
}

@inproceedings{diff-CodeQL,
author = {Froh, Fabian Niklas and Gobbi, Mat\'{\i}as Federico and Kinder, Johannes},
title = {Differential Static Analysis for Detecting Malicious Updates to Open Source Packages},
year = {2023},
isbn = {9798400702631},
publisher = {Association for Computing Machinery},
address = {New York, NY, USA},
url = {https://doi.org/10.1145/3605770.3625211},
doi = {10.1145/3605770.3625211},
booktitle = {Proceedings of the 2023 Workshop on Software Supply Chain Offensive Research and Ecosystem Defenses},
pages = {41–49},
numpages = {9},
location = {Copenhagen, Denmark},
series = {SCORED '23}
}

@inproceedings{guo2023empirical,
author = {Guo, Wenbo and Xu, Zhengzi and Liu, Chengwei and Huang, Cheng and Fang, Yong and Liu, Yang},
title = {An Empirical Study of Malicious Code In PyPI Ecosystem},
year = {2024},
isbn = {9798350329964},
publisher = {IEEE Press},
url = {https://doi.org/10.1109/ASE56229.2023.00135},
doi = {10.1109/ASE56229.2023.00135},
booktitle = {Proceedings of the 38th IEEE/ACM International Conference on Automated Software Engineering},
pages = {166–177},
numpages = {12},
location = {Echternach, Luxembourg},
series = {ASE '23}
}

@inproceedings{GENIE,
  author={Gobbi, Matías F. and Kinder, Johannes},
  booktitle={2024 IEEE Secure Development Conference (SecDev)}, 
  title={GENIE: Guarding the npm Ecosystem with Semantic Malware Detection}, 
  year={2024},
  volume={},
  number={},
  pages={117-128},
  doi={10.1109/SecDev61143.2024.00017}
}

@misc{microsoft_ossgadget,
  author       = {Microsoft},
  title        = {OSSGadget: Collection of Tools for Analyzing Open Source Packages},
  howpublished = {\url{https://github.com/microsoft/OSSGadget}},
  year         = {2020},
  note         = {Accessed: 2025-07-08}
}

@inproceedings{ohm2022feasibility,
author = {Ohm, Marc and Boes, Felix and Bungartz, Christian and Meier, Michael},
title = {On the Feasibility of Supervised Machine Learning for the Detection of Malicious Software Packages},
year = {2022},
isbn = {9781450396707},
publisher = {Association for Computing Machinery},
address = {New York, NY, USA},
url = {https://doi.org/10.1145/3538969.3544415},
doi = {10.1145/3538969.3544415},
booktitle = {Proceedings of the 17th International Conference on Availability, Reliability and Security},
articleno = {127},
numpages = {10},
location = {Vienna, Austria},
series = {ARES '22}
}

@inproceedings{ohm2020backstabber,
author = {Ohm, Marc and Plate, Henrik and Sykosch, Arnold and Meier, Michael},
title = {Backstabber’s Knife Collection: A Review of Open Source Software Supply Chain Attacks},
year = {2020},
isbn = {978-3-030-52682-5},
publisher = {Springer-Verlag},
address = {Berlin, Heidelberg},
url = {https://doi.org/10.1007/978-3-030-52683-2_2},
doi = {10.1007/978-3-030-52683-2_2},
booktitle = {Detection of Intrusions and Malware, and Vulnerability Assessment: 17th International Conference, DIMVA 2020, Lisbon, Portugal, June 24–26, 2020, Proceedings},
pages = {23–43},
numpages = {21},
location = {Lisbon, Portugal}
}

@misc{packj,
  author       = {ossillate-inc},
  title        = {Packj flags malicious/risky open-source packages},
  howpublished = {\url{https://github.com/ossillate-inc/packj}},
  year         = {2022},
  note         = {Accessed: 2025-4-16}
}

@inproceedings{huang2024donapi,
author = {Huang, Cheng and Wang, Nannan and Wang, Ziyan and Sun, Siqi and Li, Lingzi and Chen, Junren and Zhao, Qianchong and Han, Jiaxuan and Yang, Zhen and Shi, Lei},
title = {DONAPI: malicious NPM packages detector using behavior sequence knowledge mapping},
year = {2024},
isbn = {978-1-939133-44-1},
publisher = {USENIX Association},
address = {USA},
booktitle = {Proceedings of the 33rd USENIX Conference on Security Symposium},
articleno = {211},
numpages = {18},
location = {Philadelphia, PA, USA},
series = {SEC '24}
}

@inproceedings{scalco2022feasibility,
author = {Scalco, Simone and Paramitha, Ranindya and Vu, Duc-Ly and Massacci, Fabio},
title = {On the feasibility of detecting injections in malicious npm packages},
year = {2022},
isbn = {9781450396707},
publisher = {Association for Computing Machinery},
address = {New York, NY, USA},
url = {https://doi.org/10.1145/3538969.3543815},
doi = {10.1145/3538969.3543815},
booktitle = {Proceedings of the 17th International Conference on Availability, Reliability and Security},
articleno = {115},
numpages = {8},
location = {Vienna, Austria},
series = {ARES '22}
}

@Misc{securelist_lofylife,
  title = {LofyLife: malicious npm packages steal Discord tokens and bank card data},
  author = {Securelist},
  howpublished = {\url{https://securelist.com/lofylife-malicious-npm-packages/107014/}},
  year = {2022},
  note = {Accessed: 2025-07-08}
}

@inproceedings{sejfia2022practical,
author = {Sejfia, Adriana and Sch\"{a}fer, Max},
title = {Practical automated detection of malicious npm packages},
year = {2022},
isbn = {9781450392211},
publisher = {Association for Computing Machinery},
address = {New York, NY, USA},
url = {https://doi.org/10.1145/3510003.3510104},
doi = {10.1145/3510003.3510104},
booktitle = {Proceedings of the 44th International Conference on Software Engineering},
pages = {1681–1692},
numpages = {12},
location = {Pittsburgh, Pennsylvania},
series = {ICSE '22}
}

@Misc{brilworks_nodejs_stats,
  title = {50+ Node.js Statistics Covering Usage, Adoption, and Performance},
  author = {Vikas Singh},
  howpublished = {\url{https://www.brilworks.com/blog/nodejs-usage-statistics/}},
  year = {2025},
  note = {Accessed: 2025-07-08}
}

@Misc{socket_npm_retrospective,
  title = {npm in Review: A 2023 Retrospective on Growth, Security, and Quirky Facts},
  author = {Philipp Burckhardt},
  howpublished = {\url{https://socket.dev/blog/2023-npm-retrospective}},
  year = {2024},
  note = {Accessed: 2025-07-08}
}

@Misc{bacancy_nodejs_stats,
  title = {Node.js Statistics: The Updated Guide on Node.js Usage and Trends},
  author = {Ritwik Verma},
  howpublished = {\url{https://www.bacancytechnology.com/blog/nodejs-statistics}},
  year = {2026},
  note = {Accessed: 2026-03-10}
}

@inproceedings{yu2024maltracker,
author = {Yu, Zeliang and Wen, Ming and Guo, Xiaochen and Jin, Hai},
title = {Maltracker: A Fine-Grained NPM Malware Tracker Copiloted by LLM-Enhanced Dataset},
year = {2024},
isbn = {9798400706127},
publisher = {Association for Computing Machinery},
address = {New York, NY, USA},
url = {https://doi.org/10.1145/3650212.3680397},
doi = {10.1145/3650212.3680397},
booktitle = {Proceedings of the 33rd ACM SIGSOFT International Symposium on Software Testing and Analysis},
pages = {1759–1771},
numpages = {13},
location = {Vienna, Austria},
series = {ISSTA 2024}
}

@inproceedings{zahan2024leveraging,
author = {Zahan, Nusrat and Burckhardt, Philipp and Lysenko, Mikola and Aboukhadijeh, Feross and Williams, Laurie},
title = {Leveraging Large Language Models to Detect npm Malicious Packages},
year = {2025},
isbn = {9798331505691},
publisher = {IEEE Press},
url = {https://doi.org/10.1109/ICSE55347.2025.00146},
doi = {10.1109/ICSE55347.2025.00146},
booktitle = {Proceedings of the IEEE/ACM 47th International Conference on Software Engineering},
pages = {2625–2637},
numpages = {13},
location = {Ottawa, Ontario, Canada},
series = {ICSE '25}
}

@inproceedings{zahan2022weak,
author = {Zahan, Nusrat and Zimmermann, Thomas and Godefroid, Patrice and Murphy, Brendan and Maddila, Chandra and Williams, Laurie},
title = {What are weak links in the npm supply chain?},
year = {2022},
isbn = {9781450392266},
publisher = {Association for Computing Machinery},
address = {New York, NY, USA},
url = {https://doi.org/10.1145/3510457.3513044},
doi = {10.1145/3510457.3513044},
booktitle = {Proceedings of the 44th International Conference on Software Engineering: Software Engineering in Practice},
pages = {331–340},
numpages = {10},
location = {Pittsburgh, Pennsylvania},
series = {ICSE-SEIP '22}
}

@article{zhang2025killing,
author = {Zhang, Junan and Huang, Kaifeng and Huang, Yiheng and Chen, Bihuan and Wang, Ruisi and Wang, Chong and Peng, Xin},
title = {Killing Two Birds with One Stone: Malicious Package Detection in NPM and PyPI using a Single Model of Malicious Behavior Sequence},
year = {2025},
issue_date = {May 2025},
publisher = {Association for Computing Machinery},
address = {New York, NY, USA},
volume = {34},
number = {4},
issn = {1049-331X},
url = {https://doi.org/10.1145/3705304},
doi = {10.1145/3705304},
journal = {ACM Trans. Softw. Eng. Methodol.},
month = apr,
articleno = {104},
numpages = {28}
}

@inproceedings{zhang2024maldet,
  author={Zhang, Yu and Qu, Haipeng and Ying, Lingyun and Wang, Linghui},
  booktitle={2024 IEEE 23rd International Conference on Trust, Security and Privacy in Computing and Communications (TrustCom)}, 
  title={Maldet: An Automated Malicious npm Package Detector Based on Behavior Characteristics and Attack Vectors}, 
  year={2024},
  volume={},
  number={},
  pages={1942-1947},
  doi={10.1109/TrustCom63139.2024.00269}
}

@inproceedings{zimmermann2019small,
author = {Zimmermann, Markus and Staicu, Cristian-Alexandru and Tenny, Cam and Pradel, Michael},
title = {Small World with High Risks: A Study of Security Threats in the npm Ecosystem},
year = {2019},
isbn = {9781939133069},
publisher = {USENIX Association},
address = {USA},
booktitle = {Proceedings of the 28th USENIX Conference on Security Symposium},
pages = {995–1010},
numpages = {16},
location = {Santa Clara, CA, USA},
series = {SEC'19}
}

@inproceedings{garrett2019detecting,
author = {Garrett, Kalil and Ferreira, Gabriel and Jia, Limin and Sunshine, Joshua and K\"{a}stner, Christian},
title = {Detecting suspicious package updates},
year = {2019},
publisher = {IEEE Press},
url = {https://doi.org/10.1109/ICSE-NIER.2019.00012},
doi = {10.1109/ICSE-NIER.2019.00012},
booktitle = {Proceedings of the 41st International Conference on Software Engineering: New Ideas and Emerging Results},
pages = {13–16},
numpages = {4},
location = {Montreal, Quebec, Canada},
series = {ICSE-NIER '19}
}

@Misc{our_website,
  title = {Code and Dataset},
  howpublished = {\url{https://doi.org/10.6084/m9.figshare.31869370}},
  year = {2026},
  note = {Accessed: 2026-03-27}
}

@INPROCEEDINGS{ferreira2021containing,
  author={Ferreira, Gabriel and Jia, Limin and Sunshine, Joshua and Kästner, Christian},
  booktitle={2021 IEEE/ACM 43rd International Conference on Software Engineering (ICSE)}, 
  title={Containing Malicious Package Updates in npm with a Lightweight Permission System}, 
  year={2021},
  volume={},
  number={},
  pages={1334-1346},
  doi={10.1109/ICSE43902.2021.00121}
}

@ARTICLE{11037372,
  author={Wang, Jian and Li, Zhen and Qu, Jixiang and Zou, Deqing and Xu, Shouhuai and Xu, Ziteng and Wang, Zhenwei and Jin, Hai},
  journal={IEEE Transactions on Information Forensics and Security}, 
  title={MalPacDetector: An LLM-Based Malicious NPM Package Detector}, 
  year={2025},
  volume={20},
  number={},
  pages={6279-6291},
  doi={10.1109/TIFS.2025.3580336}
}

@inproceedings{ladisa2023sok,
  author={Ladisa, Piergiorgio and Plate, Henrik and Martinez, Matias and Barais, Olivier},
  booktitle={2023 IEEE Symposium on Security and Privacy (SP)}, 
  title={SoK: Taxonomy of Attacks on Open-Source Software Supply Chains}, 
  year={2023},
  volume={},
  number={},
  pages={1509-1526},
  doi={10.1109/SP46215.2023.10179304}
}

@inproceedings{neupane2023beyond,
author = {Neupane, Shradha and Holmes, Grant and Wyss, Elizabeth and Davidson, Drew and De Carli, Lorenzo},
title = {Beyond typosquatting: an in-depth look at package confusion},
year = {2023},
isbn = {978-1-939133-37-3},
publisher = {USENIX Association},
address = {USA},
booktitle = {Proceedings of the 32nd USENIX Conference on Security Symposium},
articleno = {193},
numpages = {18},
location = {Anaheim, CA, USA},
series = {SEC '23}
}

@inproceedings{taylor2020defending,
author = {Taylor, Matthew and Vaidya, Ruturaj and Davidson, Drew and De Carli, Lorenzo and Rastogi, Vaibhav},
title = {Defending Against Package Typosquatting},
year = {2020},
isbn = {978-3-030-65744-4},
publisher = {Springer-Verlag},
address = {Berlin, Heidelberg},
url = {https://doi.org/10.1007/978-3-030-65745-1_7},
doi = {10.1007/978-3-030-65745-1_7},
booktitle = {Network and System Security: 14th International Conference, NSS 2020, Melbourne, VIC, Australia, November 25–27, 2020, Proceedings},
pages = {112–131},
numpages = {20},
location = {Melbourne, VIC, Australia}
}

@inproceedings{vu2020typosquatting,
  author={Vu, Duc-Ly and Pashchenko, Ivan and Massacci, Fabio and Plate, Henrik and Sabetta, Antonino},
  booktitle={2020 IEEE European Symposium on Security and Privacy Workshops (EuroS\&PW)}, 
  title={Typosquatting and Combosquatting Attacks on the Python Ecosystem}, 
  year={2020},
  pages={509-514},
  doi={10.1109/EuroSPW51379.2020.00074}
}

@article{abdalkareem2020impact,
  title={On the impact of using trivial packages: An empirical case study on npm and pypi},
  author={Abdalkareem, Rabe and Oda, Vinicius and Mujahid, Suhaib and Shihab, Emad},
  journal={Empirical Software Engineering},
  volume={25},
  number={2},
  pages={1168--1204},
  year={2020},
  doi = {10.1007/s10664-019-09792-9},
  publisher={Springer}
}

@article{cao2022towards,
author = {Cao, Yulu and Chen, Lin and Ma, Wanwangying and Li, Yanhui and Zhou, Yuming and Wang, Linzhang},
title = {Towards Better Dependency Management: A First Look at Dependency Smells in Python Projects},
year = {2023},
issue_date = {April 2023},
publisher = {IEEE Press},
volume = {49},
number = {4},
issn = {0098-5589},
url = {https://doi.org/10.1109/TSE.2022.3191353},
doi = {10.1109/TSE.2022.3191353},
journal = {IEEE Trans. Softw. Eng.},
month = apr,
pages = {1741–1765},
numpages = {25}
}

@Misc{eslint_postmortem_2018,
  title = {Postmortem for Malicious Packages Published on July 12th, 2018},
  author = {{ESLint Team}},
  howpublished = {\url{https://eslint.org/blog/2018/07/postmortem-for-malicious-package-publishes/}},
  year = {2018},
  note = {Accessed: 2025-07-08}
}

@inproceedings{latendresse2022not,
author = {Latendresse, Jasmine and Mujahid, Suhaib and Costa, Diego Elias and Shihab, Emad},
title = {Not All Dependencies are Equal: An Empirical Study on Production Dependencies in NPM},
year = {2023},
isbn = {9781450394758},
publisher = {Association for Computing Machinery},
address = {New York, NY, USA},
url = {https://doi.org/10.1145/3551349.3556896},
doi = {10.1145/3551349.3556896},
booktitle = {Proceedings of the 37th IEEE/ACM International Conference on Automated Software Engineering},
articleno = {73},
numpages = {12},
location = {Rochester, MI, USA},
series = {ASE '22}
}

@inproceedings{wyss2025evaluating,
  title={Evaluating LLM-Based Detection of Malicious Package Updates in npm},
  author={Wyss, Elizabeth and Tassio, Dominic and De Carli, Lorenzo and Davidson, Drew},
  booktitle={2025 28th International Symposium on Research in Attacks, Intrusions and Defenses (RAID)},
  pages={678--692},
  year={2025},
  doi = {10.1109/RAID67961.2025.00047},
  organization={IEEE}
}

@inproceedings{vu2023bad,
author = {Vu, Duc-Ly and Newman, Zachary and Meyers, John Speed},
title = {Bad Snakes: Understanding and Improving Python Package Index Malware Scanning},
year = {2023},
isbn = {9781665457019},
publisher = {IEEE Press},
url = {https://doi.org/10.1109/ICSE48619.2023.00052},
doi = {10.1109/ICSE48619.2023.00052},
booktitle = {Proceedings of the 45th International Conference on Software Engineering},
pages = {499–511},
numpages = {13},
location = {Melbourne, Victoria, Australia},
series = {ICSE '23}
}

@misc{darkreading2023false,
  author       = {Robert Lemos},
  title        = {One-Third of Popular PyPI Packages Mistakenly Flagged as Malicious},
  howpublished = {\url{https://www.darkreading.com/application-security/one-third-pypi-packages-mistakenly-flagged-malicious}},
  year         = {2022},
  note         = {Accessed: 2024-12-27}
}

@inproceedings{reimers2019sentence,
  title={Sentence-bert: Sentence embeddings using siamese bert-networks},
  author={Reimers, Nils and Gurevych, Iryna},
  booktitle={Proceedings of the 2019 conference on empirical methods in natural language processing and the 9th international joint conference on natural language processing (EMNLP-IJCNLP)},
  pages={3982--3992},
  doi = {10.18653/v1/D19-1410},
  year={2019}
}

@ARTICLE{mcqueen1967some,
  author={Lloyd, S.},
  journal={IEEE Transactions on Information Theory}, 
  title={Least squares quantization in PCM}, 
  year={1982},
  volume={28},
  number={2},
  pages={129-137},
  doi={10.1109/TIT.1982.1056489}
}

@Misc{tree_sitter,
  title = {Tree-sitter: An incremental parsing system for programming tools},
  author = {Tree-sitter},
  howpublished = {\url{https://github.com/tree-sitter/tree-sitter}},
  year = {2025},
  note = {Accessed: 2025-07-08}
}

@Misc{phylum_npm_security_holding,
  title = {NPM Security Holding},
  author = {Phylum},
  howpublished = {\url{https://docs.phylum.io/analytics/npm_security_holding}},
  year = {2024},
  note = {Accessed: 2025-07-08}
}

@inproceedings{intelguard_www2026,
author = {Guo, Wenbo and Song, Shiwen and Guo, Jiaxun and Xu, Zhengzi and Liu, Chengwei and Ou, Haoran and Ge, Mengmeng and Liu, Yang},
title = {Bridging Expert Reasoning and LLM Detection: A Knowledge-Driven Framework for Malicious Packages},
year = {2026},
isbn = {9798400723070},
publisher = {Association for Computing Machinery},
address = {New York, NY, USA},
url = {https://doi.org/10.1145/3774904.3792083},
doi = {10.1145/3774904.3792083},
booktitle = {Proceedings of the ACM Web Conference 2026},
pages = {3554–3565},
numpages = {12},
location = {United Arab Emirates},
series = {WWW '26}
}

@inproceedings{profmal_ase2025,
  author={Huang, Yiheng and Zheng, Wen and Wu, Susheng and Chen, Bihuan and Lu, You and Zhou, Zhuotong and Cao, Yiheng and Li, Xiaoyu and Peng, Xin},
  booktitle={2025 40th IEEE/ACM International Conference on Automated Software Engineering (ASE)}, 
  title={ProfMal: Detecting Malicious NPM Packages by the Synergy between Static and Dynamic Analysis}, 
  year={2025},
  volume={},
  number={},
  pages={419-431},
  doi={10.1109/ASE63991.2025.00042}
}

@ARTICLE{emphunter_tse2025,
  author={Liang, Wentao and Ling, Xiang and Zhao, Chen and Wu, Jingzheng and Luo, Tianyue and Wu, Yanjun},
  journal={IEEE Transactions on Software Engineering}, 
  title={Detecting Malicious Packages in PyPI and NPM by Clustering Installation Scripts}, 
  year={2026},
  volume={52},
  number={1},
  pages={36-53},
  doi={10.1109/TSE.2025.3618952}
}

@inproceedings{zhou2025analysis,
  author={Zhou, Xiaoyan and Zhang, Ying and Niu, Wenjia and Liu, Jiqiang and Wang, Haining and Li, Qiang},
  booktitle={2025 55th Annual IEEE/IFIP International Conference on Dependable Systems and Networks (DSN)}, 
  title={An Analysis of Malicious Packages in Open-Source Software in the Wild}, 
  year={2025},
  volume={},
  number={},
  pages={371-385},
  doi={10.1109/DSN64029.2025.00045}
}

@inproceedings{zhao2024models,
  author={Zhao, Jian and Wang, Shenao and Zhao, Yanjie and Hou, Xinyi and Wang, Kailong and Gao, Peiming and Zhang, Yuanchao and Wei, Chen and Wang, Haoyu},
  booktitle={2024 39th IEEE/ACM International Conference on Automated Software Engineering (ASE)}, 
  title={Models Are Codes: Towards Measuring Malicious Code Poisoning Attacks on Pre-trained Model Hubs}, 
  year={2024},
  volume={},
  number={},
  pages={2087-2098},
  doi = {10.1145/3691620.3695271}
}

\end{document}